\documentclass[3p]{elsarticle} %review=doublespace preprint=single 5p=2 column
\usepackage[hyphens]{url}

  \journal{Faculty of Economic Sciences Working Papers series, University of Warsaw} % Sets Journal name

\usepackage{graphicx}
\usepackage[T1]{fontenc}
\usepackage{lmodern}
\usepackage{amssymb,amsmath}
\usepackage{lineno} % add
\usepackage{ifxetex,ifluatex}
\usepackage{fixltx2e} % provides \textsubscript
\IfFileExists{upquote.sty}{\usepackage{upquote}}{}
\ifnum 0\ifxetex 1\fi\ifluatex 1\fi=0 % if pdftex
  \usepackage[utf8]{inputenc}
\else % if luatex or xelatex
  \usepackage{fontspec}
  \ifxetex
    \usepackage{xltxtra,xunicode}
  \fi
  \defaultfontfeatures{Mapping=tex-text,Scale=MatchLowercase}
  
\fi
\IfFileExists{microtype.sty}{\usepackage{microtype}}{}
\usepackage[]{natbib}
\ifxetex
  \usepackage[setpagesize=false, % page size defined by xetex
              unicode=false, % unicode breaks when used with xetex
              xetex]{hyperref}
\else
  \usepackage[unicode=true]{hyperref}
\fi
\hypersetup{breaklinks=true,
            bookmarks=true,
            pdfauthor={},
            pdftitle={},
            colorlinks=true,
            urlcolor=blue,
            linkcolor=blue,
            pdfborder={0 0 0}}

\providecommand{\tightlist}{%
  \setlength{\itemsep}{0pt}\setlength{\parskip}{0pt}}

\usepackage{longtable,booktabs,array}
\usepackage{calc} % for calculating minipage widths
\usepackage{etoolbox}
\makeatletter
\patchcmd\longtable{\par}{\if@noskipsec\mbox{}\fi\par}{}{}
\makeatother
\IfFileExists{footnotehyper.sty}{\usepackage{footnotehyper}}{\usepackage{footnote}}
\makesavenoteenv{longtable}

\usepackage{floatrow}
\usepackage{adjustbox} \usepackage{lipsum} \usepackage{xcolor} \usepackage{booktabs} \usepackage{hyperref} \usepackage{tabu} \usepackage{soul} \usepackage{pdflscape} \usepackage{changepage} \usepackage{threeparttable} \usepackage[raggedrightboxes]{ragged2e} \usepackage{xurl}

\begin{document}

\begin{frontmatter}

  \title{Systemic risk indicator based on\\
implied and realized volatility}
    \author[WNEUW]{Paweł Sakowski%
  \fnref{1}}
   \ead{sakowski@wne.uw.edu.pl} 
    \author[NYU]{Rafał Sieradzki%
  \fnref{2}}
   \ead{rjs9362@nyu.edu} 
    \author[WNEUW]{Robert Ślepaczuk%
  \corref{cor1}%
  \fnref{3, 4, 5}}
   \ead{rslepaczuk@wne.uw.edu.pl} 
      \affiliation[WNEUW]{Quantitative Finance Research Group, Department of Quantitative Finance, University of Warsaw, Faculty of Economic Sciences, ul. Dluga 44-50, 00-241, Warsaw, Poland}
    \affiliation[NYU]{New York University Stern School of Business; Cracow University of Economics}
    \cortext[cor1]{Corresponding author}
    \fntext[1]{ORCID: https://orcid.org/0000-0003-3384-3795}
    \fntext[2]{ORCID: https://orcid.org/0000-0002-4702-7716}
    \fntext[3]{ORCID: https://orcid.org/0000-0001-5227-2014}
    \fntext[4]{This document is the result of the research project funded by the IDUB program BOB-IDUB-622-187/2022 at the University of Warsaw}
    \fntext[5]{We want to thank Linda Allen from Baruch College for providing us with CATFIN data, and Viral Acharya and Rob Capellini from the NYU Stern School of Business and NYU Stern Volatility and Risk Institute for sending us the sRisk data on a single constituent level. We have benefited from discussions with Tobias Adrian, Markus Brunnenmeier, Ruggero Japelli, Yi Cao, Zexun Chen. We would also like to thank participants of the 49th Eastern Economic Conference in New York (February 2023), the 33rd Quantitative Finance Research Group and Data Science Lab Research Seminar at the University of Warsaw (April 2023), the MSBE research seminar at the University of Edinburgh, School of Business (April 2023), the 43rd International Symposium on Forecasting at the University of Virginia Darden School of Business, Charlottesville, VA, USA (June 2023), and the 16th edition of the International Risk Management Conference Florence, Italy (July 2023) for their inspiring comments and insightful discussions.}
  
  \begin{abstract}
  We propose a new measure of systemic risk to analyze the impact of the major financial market turmoils in the stock markets from 2000 to 2023 in the USA, Europe, Brazil, and Japan. Our Implied Volatility Realized Volatility Systemic Risk Indicator (IVRVSRI) shows that the reaction of stock markets varies across different geographical locations and the persistence of the shocks depends on the historical volatility and long-term average volatility level in a given market. The methodology applied is based on the logic that the simpler is always better than the more complex if it leads to the same results. Such an approach significantly limits model risk and substantially decreases computational burden. Robustness checks show that IVRVSRI is a precise and valid measure of the current systemic risk in the stock markets. Moreover, it can be used for other types of assets and high-frequency data. The forecasting ability of various SRIs (including CATFIN, CISS, IVRVSRI, SRISK, and Cleveland FED) with regard to weekly returns of S\&P 500 index is evaluated based on the simple linear, quasi-quantile, and quantile regressions. We show that IVRVSRI has the strongest predicting power among them.
  \end{abstract}
  
 \end{frontmatter}

\hypertarget{introduction}{%
\section{Introduction}\label{introduction}}

\label{sec:Introduction}

The magnitude and the speed of the contagion of the financial market turmoils is the main point of interest in numerous studies. This topic is of special importance because the reactions of the financial markets to any existing or forthcoming crisis are fast, and it is hard to identify them on time based on the real economic measures, as they are announced with a delay. The main aim of this paper is to analyze and compare the systemic impact of the major financial market turmoils in the equity markets in the USA, Europe, Brazil, and Japan from 2000 to 2023. For this purpose, we construct an indicator based on implied and realized volatility measures (IV and RV, respectively) for each market, which are easily available to all market participants. Moreover, we construct a general indicator at the worldwide level. Our partial motivation to undertake this study is to show that such Systemic Risk Indicators can be constructed from simple metrics, and there is no need to use any sophisticated risk models for this purpose (\citet{caporin2021traffic}). In other words, we want to show that the model risk can be significantly reduced while the results are similar to the ones obtained by the use of much more complex tools. We set four research hypotheses:

\begin{itemize}
\tightlist
\item
  RH1: \emph{It is possible to construct a robust Systemic Risk Indicator based on the well-known concepts of realized and implied volatility measures}.
\item
  RH2: \emph{The indication of the proposed Systemic Risk Indicator depends on the geographical location of a given equity market}.
\item
  RH3: \emph{The robustness of the proposed Systemic Risk Indicator depends on various parameters selected: the memory parameter for RV, time to expiration for IV, the percentile selected for the risk map, the length of the history selected for the calculation of percentile in case of risk map}.
\item
  RH4 \emph{IVRVSRI has the highest forecasting ability of S\&P500 index amongst other benchmark SRIs, especially in the moments of systemic risk}
\end{itemize}

The robustness of proposed systemic risk measure is particularly important, as in many studies (e.g. \citet{caporin2021traffic}, \citet{hollo2012ciss}) researchers do not consider extent to which the initial parameters of the model affect the final results, especially those regarding the speed of reaction to unexpected market turmoils. We check the sensitivity of the proposed Systemic Risk Indicator to the change of the selected parameters like: the memory parameter for the realized volatility (RV), time to expiration for the implied volatility (IV), the percentile selected for the risk map, and the length of the history selected for the calculation of percentile in case of the risk map.

Systemic risk refers to the risk of the collapse of an entire financial system, as opposed to the risk associated with any individual entity, which is a credit default risk. One of the main distinguishing features of systemic risk is that an idiosyncratic event affecting one or a group of market entities is exacerbated by interlinkages and interdependencies in a system, leading to a domino effect that can potentially bring down the entire system. The fragility of the system as a whole is being built over time, and it ``only'' materializes in times of crisis. One reason is a comparable business model among market participants that leads to accumulating similar assets on a balance sheet and comparable investment strategies that resemble herd behavior. In fact, those entities, although legally separated, can be considered as one large entity from a systemic point of view.

The recent collapses of Silicon Valley Bank and Signature Bank, and the abrupt nature of the Covid-19 pandemic, suggest the timeliness of the indicators is one of the key characteristics of a systemic risk indicator. One may claim that relying on the measures primarily based on the market variables, basically the prices of the financial instruments, may be potentially misleading as they may generate false positive signals. On the other hand, measures that are based primarily on accounting-based data are slow in reacting to potential problems in the financial system, as they are available with a delay. Therefore, we argue that it is better to get a signal of a potential problem that sometimes may be false that to get it when the crisis has already started. Most of the research recognizes that problem and tries to combine market and accounting data (See the literature review for details).

In general, the function of the market-based variables is to detect potential crises in a timely manner, and the accounting-based measures serve to identify systematically important institutions, whose collapse may create spill-over effects in the system. This approach may be applied by the market regulators that can monitor more closely entities that are systematically important, and try to introduce regulatory solutions to lower their impact on the system. These entities also have access to more data on individual institutions and also collect them earlier than market participants. Although, we agree that this approach is a good way to identify the ``too-big-to-fail'' institutions. On the other hand, we argue that one of the drawbacks of this approach is that combining both types of data leads to higher model risk and is computationally more intensive than using only market-based variables. Moreover, it seems to be hard to detect the interconnectedness in the market due to its complex and dynamic nature and therefore classifying ``too-interconnected-to-fail'' is a difficult task\footnote{ Some systemic risk measures, like $\Delta$CoVaR, try to capture the potential for the spreading of financial distress across institutions by gauging spill-over by observing the tail comovement using the VaR approach}. At the same time, the aforementioned collapses of the SVB and Signature Bank show that some risk built-up in the system we were not aware of and they were not taken into account by the existing models\footnote{Those banks had bonds on their balance sheets which were “held to maturity”. When the customers started to withdraw their deposits, banks had to sell those bonds in the market at much lower prices than they were reported on the balance sheet, leading to huge losses.}.

In this work, we focus on one part of the systemic risks, which is the timely identification of the potential crisis by market participants and not only by the regulators. We also claim that a ``wide market'' may have superior knowledge about systemic risk, and to some extent, their coordinated actions can trigger a systemic event\footnote{There is anecdotal evidence that some depositors withdrew all their funds from the SVB two days before its collapse. It seems, that presumably the involuntarily coordinated action of a group of investors to take out their deposits from that bank in a short stretch of time was the trigger of the bank’s collapse. At the same time, one may assume that those investors who were very convinced that the SVB was going to have serious problems were short-selling its stocks or going long deep out-of-the-money options on its stocks, further exacerbating the problems of the bank and finally leading to its collapse.}.

The structure of this paper is as follows. The second section presents a literature review. The third section describes Data and Methodology. The fourth section presents the Results, and the fifth one includes Conclusions.

\hypertarget{literature-review-and-classification-the-selected-systemic-risk-indicators}{%
\section{Literature review and classification the selected systemic risk indicators}\label{literature-review-and-classification-the-selected-systemic-risk-indicators}}

\label{sec:LiteratureReview_and_classification}

\hypertarget{literature-review}{%
\subsection{Literature review}\label{literature-review}}

\label{sec:LiteratureReview}

The major approach in the literature to measure systemic risk is based either on market data or a mix of market and balance sheet data. Those combined risk indicators use i.a. such metrics as VaR and CoVaR. The results obtained for one country, market segment, or economic sector are aggregated to get a general measure of systemic risk. In general, various methods yield similar results as in \citet{engle2018much}, \citet{brownlees2017srisk}, \citet{acharya2017measuring}, \citet{bisias2012survey} or \citet{caporin2021traffic}.

One of the first attempts focusing on systemic risk was \citet{brimmer1989distinguished} who reminded the last resort lending function of the central bank, which has digressed from its overall strategy of monetary control to also undertake a tactical rescue of individual banks and segments of the financial market. \citet{de2000systemic} developed a broad concept of systemic risk, the basic economic concept for the understanding of financial crises. They claimed that any such concept must integrate systemic events in banking and financial markets as well as in the related payment and settlement systems. At the heart of systemic risk are contagion effects, and various forms of external effects. The concept also includes simultaneous financial instabilities following aggregate shocks. They surveyed the quantitative literature on systemic risk, which was evolving swiftly in the last couple of years.

\citet{bisias2012survey} point out that systemic risk is a multifaceted problem in an ever-changing financial environment, any single definition is likely to fall short and may create a false sense of security as financial markets evolve in ways that escape the scrutiny of any one-dimensional perspective. They provide an overview of over 30 indicators of systemic risk in the literature, chosen to address key issues in measuring systemic risk and its management. The measures are grouped into six various categories including: macroeconomic, granular foundations and network, forward-looking risk, stress-test, cross-sectional, illiquidity and finally insolvency measures. They analyze them from the supervisory, research, and data perspectives, and present concise definitions of each risk measure. At the same time, they point out that the system to be evaluated is highly complex, and the metrics considered were largely untested outside the GFC crisis. Indeed, some of the conceptual frameworks that they reviewed were still in their infancy and had yet to be applied.

\citet{schwarcz2008systemic} agreed that governments and international organizations worried increasingly about systemic risk, under which the world's financial system could have collapsed like a row of dominoes. There is widespread confusion, though, about the causes and, to some extent, even the definition of systemic risk, and uncertainty about how to control it. His paper offers a conceptual framework for examining what risks are truly ``systemic,'' what causes those risks, and how, if at all, those risks should be regulated. Scholars historically have tended to think of systemic risk primarily in terms of financial institutions such as banks. However, with the growth of disintermediation, in which companies can access capital-market funding without going through banks or other intermediary institutions\textbackslash footnoote\{In the US more than 50\% of funding of non-financial corporations comes from equity and bond issuance.\}, greater focus should be devoted to financial markets and the relationship between markets and institutions. This perspective reveals that systemic risk results from a type of tragedy of the commons in which market participants lack sufficient incentives, and absence of the regulation to limit risk-taking in order to reduce the systemic danger to others.

In this light, \citet{acharya2009theory} models systemic risk is modeled as the endogenously chosen correlation of returns on assets held by banks. The limited liability of banks and the presence of a negative externality of one bank's failure on the health of other banks give rise to a systemic risk-shifting incentive where all banks undertake correlated investments, thereby increasing economy-wide aggregate risk. Regulatory mechanisms such as bank closure policy and capital adequacy requirements that are commonly based only on a bank's own risk fail to mitigate aggregate risk-shifting incentives, and can, in fact, accentuate systemic risk. Prudential regulation is shown to operate at a collective level, regulating each bank as a function of both its joint (correlated) risk with other banks as well as its individual (bank-specific) risk.

\citet{brownlees2017srisk} introduce SRISK to measure the systemic risk contribution of a financial firm. SRISK captures the capital shortfall of a firm conditional on a severe market decline and is a function of its size, leverage and risk. They use the measure to study the top financial institutions in the recent financial crisis. SRISK delivers useful rankings of systemic institutions at various stages of the crisis and identifies Fannie Mae, Freddie Mac, Morgan Stanley, Bear Stearns, and Lehman Brothers as the top contributors as early as 2005-Q1. Moreover, aggregate SRISK provides early warning signals of distress in indicators of real activity.

The \(\Delta\)CoVaR method proposed by \citet{101257aer20120555} estimates the systemic risk of a financial system conditional on institutions being in distress based on publicly traded financial institutions. They define an institution's contribution to systemic risk as the difference between \(\Delta\)CoVaR conditional on the institution being in distress and \(\Delta\)CoVaR in the median state of the institution. They quantify the extent to which characteristics such as leverage, size, and maturity mismatch predict systemic risk contribution.

\citet{romer2017new} examine the aftermath of the postwar financial crises in advanced countries. Through the construction of a semiannual series of financial distress in 24 OECD countries for the period 1967--2012. The series is based on assessments of the health of countries' financial systems from and classifies financial distress on a relatively fine scale. They find that the average decline in output following a financial crisis is statistically significant and persistent, but only moderate in size. More importantly, the average decline is sensitive to the specification and sample, and that the aftermath of the crises is highly variable across major episodes. Following this research, \citet{engle2018much}, using a crisis severity variable constructed by \citet{romer2017new}, estimated a Tobit model for 23 developed economies. They developed a probability of crisis measure and SRISK capacity measure from the Tobit estimates. These indicators reveal an important global externality whereby the risk of a crisis in one country is strongly influenced by the undercapitalization of the rest of the world.

\citet{acharya2017measuring} present an economic model of systemic risk in which undercapitalization of the financial sector as a whole is assumed to harm the real economy, leading to a systemic risk externality. Each financial institution's contribution to systemic risk can be measured as its systemic expected shortfall (SES), that is, its propensity to be undercapitalized when the system as a whole is undercapitalized.

The research by \citet{WANG2022102361} addresses the measurement of the systemic risk contribution (SRC) of country-level stock markets to understand the rise of extreme risks worldwide to prevent potential financial crises. The proposed measure of SRC is based on quantifying tail risk propagation's domino effect using CoVaR and the cascading failure network model. While CoVaR captures the tail dependency structure among stock markets, the cascading failure network model captures the nonlinear dynamic characteristics of tail risk contagion to mimic tail risk propagation. The validity test demonstrated that this method outperforms seven classic methods as it helps early warning of global financial crises and correlates to many systemic risk determinants, e.g., market liquidity, leverage, inflation. The results highlight that considering tail risk contagion's dynamic characteristics helps avoid underestimating SRC and supplement a ``cascading impact'' perspective to improve financial crisis prevention.

The micro-level methods have been criticized by \citet{ALLEN20161}. They base their research on the assumption that financial intermediaries including commercial banks, savings banks, investment banks, broker/dealers, insurance companies, mutual funds, etc. are special because they are fundamental to the operation of the economy. The specialness of banks is reflected in the economic damage that results when financial firms fail to operate properly. They proposed a new measure to forecast the likelihood that systemic risk-taking in the banking system as a whole, called CATFIN. It captures the tail risk of the overall banking market using VaR methodology at a 1\% level with monthly data. This early warning system should signal whether aggressive aggregate systemic risk-taking in the financial sector presages future macroeconomic declines. \citet{GAO2022102826} showed that among 19 different risk measures, CATFIN performs the best in predicting macro-level shocks.

\citet{caporin2021traffic} introduced TALIS (TrAffic LIght System for Systemic Stress) that provides a comprehensive color-based classification for grouping companies according to both the stress reaction level of the system when the company is in distress and the company's stress. level. This indicator can integrate multiple signals from the interaction between different risk metrics. Starting from specific risk indicators, companies are classified by combining two loss functions, one for the system and one for each company, evaluated over time and as a cross-section. An aggregated index is also obtained from the color-based classification of companies.

\citet{KielakSlepaczuk2020} compare different approaches to Value-at-Risk measurement based on parametric and non-parametric approaches for different portfolios of assets, including cryptocurrencies. They checked if the analyzed models accurately estimate the Value-at-Risk measure, especially in the case of assets with various returns distribution characteristics (eg. low vs.~high volatility, high vs.~moderate skewness). \citet{RePEcWarWpaper2019-12} checked which of the VaR models should be used depending on the state of the market volatility. They showed that GARCH(1,1) with standardized student's t-distribution is least affected by changes in volatility among analysed models. \citet{RePEcWarWpaper2021-10} point out that under the conditions of sudden volatility increase, such as during the global economic crisis caused by the Covid-19 pandemic, no classical VaR model worked properly even for the group of the largest market indices. In general, there is an agreement between market risk researchers that an ideal model for VaR estimation does not exist, and different models' performance strongly depends on current economic circumstances.

Some spectacular crash events, including the FTX collapse in November 2022, followed by a dramatic slump in prices of most of the cryptocurrencies triggered a question about the resiliency of this financial market segment to shocks and the potential spillover effect. In one of the latest research, \citet{JALAN2023103670} studied systemic risk in the cryptocurrency market based on the FTX collapse. Using the CATFIN measure to proxy for the systemic risk they claimed that the FTX crisis did not engender higher systemic and liquidity risks in this market compared to previous negative shocks.

Various rigorous models of bank and payment system contagion have now been developed, although a general theoretical paradigm is still missing. Direct econometric tests of bank contagion effects seem to be mainly limited to the United States. Empirical studies of the systemic risk in foreign exchange and security settlement systems appear to be non-existent. Moreover, the literature surveyed reflects the general difficulty to develop empirical tests that can make a clear distinction between contagion in the proper sense and joint crises caused by common shocks, rational revisions of depositor or investor expectations when information is asymmetric (``information-based'' contagion) and ``pure'' contagion as well as between ``efficient'' and ``inefficient'' systemic events.

Bearing in mind the huge dynamics of the recent shocks (e.g.~the Covid-19 pandemic, and the FTX collapse), we claim that the monthly data frequency (like in the case of CATFIN) is not enough to create a valid early warning indicator. At the same time, we claim that the existing indicators of systemic risk are over sophisticated and some of them require huge computing power or access to paid datasets. Therefore, there is a need to create a precise and simple indicator of systemic risk based on a publicly available date with relatively high frequency. In this study, we base on the macro-level data which is easily accessible to the general public to construct a robust systemic risk indicator. We show that our simple metrics can yield similar (or better) results than complex methods and can be computed with a relatively high-frequency using publicly available data, which is a great advantage.

\hypertarget{a-comparison-of-the-selected-systemic-risk-indicators}{%
\subsection{A comparison of the selected systemic risk indicators}\label{a-comparison-of-the-selected-systemic-risk-indicators}}

\label{sec:Classification}

Following the Cleveland Fed's commentary on the performance of their systemic risk indicator (Craig 2020), we agree that a good financial-stress indicator (we may also say a good systemic risk indicator) is reliable, timely, straightforward, valid, and ongoing. Most of the indicators miss some of those features. For example, indicators that base on the balance-sheet data are neither timely nor ongoing, as financial data is provided on a monthly basis to the regulators and it is publicly released on a quarterly basis and with a delay. This means that those indicators can be computed by market regulators with a higher frequency than by the wide public, which is a disadvantage for the market participants. Moreover, some of the indicators are complex and thus they involve a significant model risk. In other words, if two indicators perform the same, the better one is the simpler one. In Table \ref{tab:tableComparison} we provide an overview of the selected systemic risk indicators.

\begin{landscape}
\begin{table}[!ht]
\begin{adjustwidth}{+0.05in}{0in} 
\centering
\caption{\bf A comparison of the selected systemic risk indicators}
\fontsize{6}{7}\selectfont
\sffamily
\begin{tabular}{|p{0.31in}|p{0.24in}|p{0.49in}|p{0.3in}|p{3.35in}|p{0.5in}|p{0.6in}|p{0.3in}|p{0.3in}|p{0.25in}|p{0.3in}|p{0.25in}|}\hline

 \textbf{Indicator [year of publication]}   &   \textbf{Data frequency*}    &   \textbf{Scope
 /Focus}    &   \textbf{Markets}    &   \textbf{Description} & \textbf{Methodology} &   \textbf{Link to website with updated systemic risk measures data}&  \textbf{Comple\-xity of the model}  &   \textbf{Real-time / HF data}    & \textbf{Publi\-cation lag** (days)} & \textbf{Type of data}   &   \textbf{Refe\-rences}   \\ 
\hline \hline
IVRVSRI [2023] &    Daily   &   Equities    &   Global, USA, Europe, Japan, Brazil  &   The methodology is based on the combination of the information hidden in the latent process of volatility using the concept of implied and realized volatility. This indicator includes the publicly available index of implied volatility and realized volatility on the underlying equity index.  & RV estimators, IV indices based on the concept of log contract \cite{demeterfi1999more} & \url{http://qfrg.wne.uw.edu.pl/projects/IVRVSRI} & Low  &   Possible & 0 &  Public, available in real-time  & (this study)  
\\ \hline
Talis3  [2021] &    Daily   &   US Financial institutions   &   USA &   A TrAffic LIght System for Systemic Stress (TALIS-cube) provides a color-based classification for grouping financial companies according to the system’s stress reaction level when the company is in distress. TALIS3 integrates multiple signals from the interaction between different risk metrics. Starting from specific risk indicators, companies are classified by combining two loss functions: one for the system and one for each company that is evaluated over time and as a cross-section. An aggregated index is presented in the form of a color-based classification of companies.    & VaR, CoVaR, $\Delta CoVaR$, quantile regression &  &  High    &   Possible but highly time-consuming  & 0 &   Public, available in real-time & \cite{caporin2021traffic}  
\\ \hline
Systemic Expected Shortfall (SES)   [2016] &    Monthly &   Banks   &   USA &   The idea of systemic risk is rooted in the undercapitalization of the financial sector as a whole as it is assumed to harm the real economy, leading to a systemic risk externality. Each financial institution’s contribution to systemic risk can be measured as its systemic expected shortfall (SES), that is, its propensity to be undercapitalized when the system as a whole is undercapitalized. SES increases in the institution’s leverage and its marginal expected shortfall (MES), that is, its losses in the tail of the system’s loss distribution.  & VaR, ES, marginal ES (MES), systemic ES (SES) &   \url{https://vlab.stern.nyu.edu/docs/srisk} &   High &  Not possible    & 0 &   Public, available with delay    &   \cite{acharya2017measuring} 
\\ \hline
Srisk   [2017] &    Monthly &   Depositories including banks, insurers, broker-dealers &    All countries   &   Srisk measures the systemic risk contribution of a financial firm. Srisk measures the capital shortfall of a firm conditional on a severe market decline and is a function of its size, leverage, and risk. Moreover, aggregate Srisk provides early warning signals of distress in indicators of real activity.& ES, GARCH-DCC &   \url{https://vlab.stern.nyu.edu/docs/srisk} &   High &  Possible    &   0 & Public, available with delay    &   \cite{brownlees2017srisk}   
\\ \hline
CoVaR   [2009] &    Daily   &   4 financial institutions    &   USA &   CoVaR measure of the systemic risk is defined as the change in the Value at Risk of the financial system conditional on an institution being under distress relative to its median state. Such characteristics as leverage, size, maturity mismatch, and asset price booms significantly predict CoVaR. & VaR, CoVaR, ES CoES   &    &  High    &   Not possible    &  0 &  Public, available with delay    &   \cite{101257aer20120555}    
\\ \hline
Cleveland Fed’s Systemic Risk Indicator [2013] &    Daily   &   Banks   &   USA &   This indicator combines measures of balance-sheet strength, volatility, and correlation of the asset values of the major banks with the forward-looking characteristics of option prices. This method uses the concept of the distance to default, a measure developed by Merton (1974) for firms such as banks that are highly leveraged. It is based on the calculation of two measures of insolvency risk, one an average of default risk across individual banking institutions (average distance-to-default) and the other a measure of risk for a weighted portfolio of the same institutions (portfolio distance-to-default). The systemic risk indicator is the difference (spread) between the two. When the insolvency risk of the banking system as a whole rises and converges to the average insolvency risk of individual banking institutions—the narrowing of the spread—it reflects market perceptions of imminent systematic disruption of the banking system & Merton model (\cite{Merton1974OnThePricing}for equity valuation   &   \url{https://www.clevelandfed.org/indicators-and-data/systemic-risk-indicator}  &    &  Not possible    & & Public, available with a delay  &   
\cite{craig2020How}, \cite{Zambrana2013Systemic}, \cite{Merton1974OnThePricing}
\\ \hline
CATFIN  [2012] &    Monthly &   Banks   &   USA, Europe, Asia   &   CATFIN is a crosssectional tail risk measure based on equity returns for all financial firms. CATFIN can be used to forecast macroeconomic declines around six months into the future. Using out of sample estimation, conducted with US, European and Asian bank data, they derive an early warning threshold such that CATFIN levels above this threshold (denoted $\overline{CATFIN}$) signal imminent financial crisis and recession. The CATFIN measure is an aggregate macro-level estimate of systemic risk in the banking sector. However, it does not measure the contribution of each individual bank to the overall level of systemic risk. 

This measure is based on the concept of the bank "specialness" in the economy. High levels of systemic risk in the banking sector impact the macroeconomy through aggregate lending activity. A conditional asset pricing model shows that CATFIN is priced for financial and non-financial firms. &    &       &    &  Not possible &  &   Public, available with delay    &   \cite{ALLEN20161}, \cite{allen2012does} 
\\ \hline
CISS    ECB [2012] &    Weekly  &   money, bond, equity, forex exchange market and financial intermediaries &   Euro Area   &   Composite Indicator of Systemic Stress (CISS) is supposed to be a new indicator of contemporaneous stress in the financial system. The main methodological innovation of the CISS is the application of basic portfolio theory to the aggregation of five market-specific subindices created from a total of 15 individual financial stress measures. The aggregation accordingly takes into account the time-varying cross-correlations between the subindices. As a result, the CISS puts relatively more weight on situations in which stress prevails in several market segments at the same time, capturing the idea that financial stress is more systemic and thus more dangerous for the economy as a whole if financial instability
spreads more widely across the whole financial system.  &   RV, threshold VAR model, time-varying cross-correlations estimated recursively on the basis of EWMA  & \url{https://sdw.ecb.europa.eu/browseExplanation.do?node=9689686}     & High &   Not possible    & & Public and not public (from ECB), available with delay   &  \cite{hollo2012ciss}    \\ \hline
\end{tabular}

\begin{flushleft} Source: Author’s own. 
(*) Data frequency is determined by the lowest frequency of the data used in the calculation of a systemic risk indicator. Some indicators combine daily (market) and monthly data (balance sheet), and the indicators are presented on a daily basis. We claim that it is not appropriate, and in fact, such indicators are monthly. Moreover, one should bear in mind that balance sheet data is available with a delay, which further reduces the indicators’ timeness. 
(**) The maximum lag between the moment the data is gathered and the moment of using it in the final value of the indicator.
\end{flushleft}
\label{tab:tableComparison}
\end{adjustwidth}
\end{table}
\end{landscape}

\hypertarget{methodology}{%
\section{Methodology}\label{methodology}}

\label{sec:Methodology}

Our methodology is based on the combination of the information hidden in latent process of volatility using the concept of implied and realized volatility. We did it by utilizing the methodology for volatility indices based on \citet{demeterfi1999more} and \citet{cboe2003vix} and the concept of realized volatility for various frequency of data introduced by \citet{andersen2003modeling} and \citet{andersen1999realized}.

Similarly to \citet{caporin2021traffic}, we construct a dynamic historical ranking evaluating the systemic risk day by day both on the global and country level. What is more important, our methodology can be transformed and adapted in a simple way for the use with high-frequency data and so that such systemic risk indicator can monitor the risk on the real-time basis.

The general formula of the IVRVSRI consist of two component indices which are based on implied (IVSRI) and realized (RVSRI) volatility.

\hypertarget{implied-volatility---volatility-indices}{%
\subsection{Implied volatility - Volatility indices}\label{implied-volatility---volatility-indices}}

\label{sec:ImpliedVolatility}

One of the first and the widely known volatility index is the VIX index, introduced by CBOE in 2003 and recalculated backward to 1987. Its formula, based on the seminal paper of \citet{demeterfi1999more}, was described in detail in \citet{cboe2003vix} and it can be summarized by the following equation:

\begin{equation}
\label{eq:VIX}
\sigma^2 =  \frac{2}{T} \sum_{k=i}^{} \frac{\Delta K_i}{K_i^2} e^{RT} Q(K_i) - \frac{1}{T}\Big[\frac{F}{K_0}-1\Big]^2
\end{equation}
where:

\(\sigma = \mathrm{VIX}/100\),

\(T\) - time to expiration,

\(K_i\) - strike price of \(i\)-th out-of-the-money option; a call if \(K_i > K_0\) and a put if \(K_i < K_0\); both put and call if \(K_i = K_0\),

\(R\) - risk-free interest rate to expiration,

\(F\) - forward index level derived from index option prices,

\(K_0\) - first strike below the forward index level (\(F\)).

The formulas for other volatility indices used in this study (VSTOXX, VNKY, and VXEWZ) are based on the similar methodology and their details can be found in \citet{borse2005guide}, \citet{nikkei2012guide}, and \citet{cboe2003vix}.

\hypertarget{realized-volatility-measure}{%
\subsection{Realized volatility measure}\label{realized-volatility-measure}}

\label{sec:RealizedVolatility}

In the case of historical volatility measure, we use the realized volatility concept (\citet{andersen1999realized}). It is based on summation of log returns during a given period of time and annualized in order to combine it later with IV. The formula used in this paper is as follows:

\begin{equation}
\label{eq:RVi}
\mathrm{RV}_{t,i}^{1M} = \sqrt{\frac{252}{21}\sum_{k=0}^{20} r_{t-k,i}^2} = \mathrm{RVSRI}_i, \quad\quad r_{t,i} = \log\Big(\frac{P_{t,i}}{P_{t-1,i}}\Big)
\end{equation}
where \(\mathrm{RV}_{t,i}^{1M}\) is the realized volatility for \(i\)-th equity index on day \(t\) with the memory of 1 calendar month (i.e.~21 trading days), while \(P_{t,i}\) is the price of \(i\)-th equity index on day \(t\).

The memory of the realized volatility estimator was set to 21 days (trading days) in order to make it comparable with 30 calendar days in case of VIX.

\hypertarget{ivrvsri---implied-volatility-realized-volaitlity-systemic-risk-indicator}{%
\subsection{IVRVSRI - Implied Volatility Realized Volaitlity Systemic Risk Indicator}\label{ivrvsri---implied-volatility-realized-volaitlity-systemic-risk-indicator}}

\label{sec:IVRVSRI}

Our methodology has significant advantages compared to other approaches presented in the literature (\citet{caporin2021traffic}, \citet{RePEcWarWpaper2019-12}). First, IVRVSRI uses systemic risk indication based on two simple and heavily grounded amongst market participants risk measures (IV and RV). Second, we analyze various financial market turmoils from 2000 until 2023 unhiding the characteristics and severity of major market crisis during the last 23 years. Third, we construct a dynamic ranking (day by day) showing the current level of stress on the global level and additionally separately for USA, Europe, Brazil and Japan. Finally, our methodology can be simply extended by using high-frequency price data for the selected equity indices and the same frequency for volatility indices to mimic the systemic-risk on real time basis.

In order to accomplish this task we construct two component systemic risk indicators based on implied (IVSRI) and realized volatility measures (RVSRI) for each country separately and additionally on the aggregated level for all countries.

\hypertarget{implied-volatility-sri}{%
\subsubsection{Implied Volatility SRI}\label{implied-volatility-sri}}

\label{sec:IVSRI}

IVSRI is based on the separate volatility index for each country, or group of countries, and its share in the total market capitalization. The formula for IVSRI is as follows:

\begin{equation}
\label{eq:IVSRI}
\mathrm{IVSRI} = \sum_{k=1}^{N} w_i * \mathrm{IV}_i
\end{equation}
where \(N\) is the number of analyzed countries, \(\mathrm{IV}_i\) denotes the implied volatility index for the \(i\)-th country, and \(w_i\) is the weight of the given country in SRI, calculated according to:

\begin{equation}
\label{eq:weight}
w_i= \frac{\mathrm{MC}_i}{\sum_{k=1}^{N} \mathrm{MC}_i}
\end{equation}
where \(\mathrm{MC}_i\) is the market capitalization fo the given country.

Based on Table \ref{tab:MarketCap} and Equation \ref{eq:weight}, we construct weights vector \(w\) = \{77.7\%, 8.1\%, 12\%, 2.2\%\} which will be used in calculations of our risk metrics.

\hypertarget{realized-volatility-sri}{%
\subsubsection{Realized Volatility SRI}\label{realized-volatility-sri}}

\label{sec:RVSRI}

RVSRI is based on similar concept as IVSRI (section \ref{sec:IVSRI}):

\begin{equation}
\label{eq:RVSRI}
\mathrm{RVSRI} = \sum_{k=1}^{N} w_i * \mathrm{RV}_i
\end{equation}
where \(\mathrm{RV}_i\) the realized volatility index for the \(i\)-th country.

\hypertarget{ivrvsri---implied-volatility-realized-volatility-systemic-risk-indicator-on-the-country-level}{%
\subsubsection{IVRVSRI - Implied Volatility Realized Volatility Systemic Risk Indicator on the country level}\label{ivrvsri---implied-volatility-realized-volatility-systemic-risk-indicator-on-the-country-level}}

\label{sec:IVRVcSRI}

IVRVSRI can be calculated on the country level (\(\mathrm{IVRVSRI}_i\)) as the weighted sum of \(\mathrm{IV}_i\) and \(\mathrm{RV}_i\) measures for the given country and on the global level:

\begin{equation}
\label{eq:IVRVSRI_i}
\mathrm{IVRVSRI}_i = w_{IV} * \mathrm{IVSRI}_i + w_{RV} * \mathrm{RVSRI}_i
\end{equation}

\hypertarget{ivrvsri---implied-volatility-realized-volaitlity-systemic-risk-indicator-on-the-global-level-ivrvsri}{%
\subsubsection{IVRVSRI - Implied Volatility Realized Volaitlity Systemic Risk Indicator on the global level (IVRVSRI)}\label{ivrvsri---implied-volatility-realized-volaitlity-systemic-risk-indicator-on-the-global-level-ivrvsri}}

\label{sec:IVRVSRI-2}

IVRVSRI can be calculated in both ways, based on IVSRI and RVSRI on the global level (formulas \ref{eq:IVSRI} and \ref{eq:RVSRI}):

\begin{equation}
\label{eq:IVRVSRI}
\mathrm{IVRVSRI} = w_{IV} * \mathrm{IVSRI} + w_{RV} * \mathrm{RVSRI}, \quad\quad w_{IV} + w_{RV} = 1
\end{equation}
where \(w_{IV}\) is the weight of IVSRI component in IVRVSRI (equal to 50\%), and \(w_{RV}\) is the weight of RVSRI component in IVRVSRI (equal to 50\%).

Alternatively, we can calculate IVRVSRI measure based on country specific IVRVSRI (i.e.~\(\mathrm{IVRVSRI_i}\))

\begin{equation}
\label{eq:IVRVSRI-2}
\mathrm{IVRVSRI} = \sum_{k=1}^{N} w_i * \mathrm{IVRVSRI}_i
\end{equation}

The weights \(w_{IV}\) and \(w_{RV}\) are assumed to be equal, however, different different weights may also be used.

\hypertarget{dynamic-quartile-ranking-based-on-ivrvsri-dqr_ivrvsri}{%
\subsection{Dynamic quartile ranking based on IVRVSRI (DQR\_IVRVSRI)}\label{dynamic-quartile-ranking-based-on-ivrvsri-dqr_ivrvsri}}

In the next step, we construct the dynamic quartile ranking (DQR\_IVRVSRI) based on RVSRI, IVSRI, and IVRVSRI indications, both on the country and on the global level.

The DQR\_IVRVSRI on the country level is constructed based on the following steps:

\begin{enumerate}
\def\labelenumi{\arabic{enumi}.}
\tightlist
\item
  We create quartile map chart based on \(\mathrm{IVRVSRI}_i\) for each country under investigation,
\item
  This map chart on the daily level shows colored systemic risk indicator,
\item
  Colors indicate the following:

  \begin{itemize}
  \tightlist
  \item
    RED, if \(\mathrm{IVRVSRI}_i\) is in its 4th quartile based on historical indications \(\rightarrow\) VERY HIGH country-systemic risk,
  \item
    ORANGE, if \(\mathrm{IVRVSRI}_i\) is in its 3rd quartile based on historical indications \(\rightarrow\) HIGH country-systemic risk,
  \item
    LIGHT GREEN, if \(\mathrm{IVRVSRI}_i\) is in its 2nd quartile based on historical indications \(\rightarrow\) LOW country-systemic risk,
  \item
    GREEN, if \(\mathrm{IVRVSRI}_i\) is in its 1st quartile based on historical indications \(\rightarrow\) VERY LOW country-systemic risk.
  \end{itemize}
\end{enumerate}

To construct the index at the global level, we follow the same approach as for the \(\mathrm{IVRVSRI}_i\) for each country separately, however the map is constructed on the global level.

\hypertarget{benchmark-systemic-risk-indicators}{%
\subsection{Benchmark systemic risk indicators}\label{benchmark-systemic-risk-indicators}}

\label{sec:Benchmark}

From an array of systemic risk measures described in Section \ref{sec:Classification}, we select four indicators that will serve as benchmarks for the IVRVSRI. We choose the sRisk, the CATFIN, the Cleveland FED Systemic Risk Indicator, and the CISS. As this choice may seem arbitrary, it is partly based on the availability of the data. Moreover, as the selected measures use different methodologies, we want to check how their indications compare with each other and how accurate they are at predicting financial turmoils. Data for the sRisk is available for many countries, and at the global level, while the CATFIN and Cleveland FED's measures are available only for the US market, and the CISS indicator is only available for Europe.

\hypertarget{srisk}{%
\subsubsection{SRISK}\label{srisk}}

\label{sec:SRISK}

SRISK is defined as the expected capital shortfall of a financial entity conditional on a prolonged market decline. It is a function of the size of the firm, its degree of leverage, and its expected equity loss conditional on the market decline, which is called Long Run Marginal Expected Shortfall (LRMES). The SRISK calculation is analogous to the stress tests that are regularly applied to financial firms. It is done with only publicly available information, making the index widely applicable and relatively inexpensive to implement for a single entity. The measure can readily be computed using balance sheet information and an appropriate LRMES estimator. Firms with the highest SRISK are the largest contributors to the undercapitalization of the financial system in times of distress. The sum of SRISK across all firms is used as a measure of overall systemic risk in the entire financial system. It can be thought of as the total amount of capital that the government would have to provide to bail out the financial system in case of a crisis. SRISK combines market and balance sheet information in order to construct a market-based measure of financial distress, which is the expected capital shortfall of a financial firm conditional on a systemic event. SRISK depends not only on equity volatility and correlation (or other moments of the equity return distribution), but also explicitly on the size and the degree of leverage of a financial firm.

According to \citet{brownlees2017srisk}, SRISK can be calculated based on the following formulas. They start from the definition of the capital shortfall of firm \(i\) on day \(t\):

\begin{equation}
\label{eq:CapitalShortfall}
\mathrm{CS}_{it} = kA_{it}-W_{it}=kt(D_{it}+W_{it})
\end{equation}
where:

\(W_{it}\) - is the market value of equity,

\(D_{it}\) - is the book value of debt,

\(A_{it}\) - is the value of quasi asset,

\(k\) - is the prudential capital fraction (it is assumed on the level of 8\%).

Then, they note that when the capital shortfall is negative, i.e., the firm has a capital surplus, the firm functions properly. However, when this quantity is positive, the firm experiences distress. They add, that the main focus will be put on the prediction of the capital shortfall of a financial entity in case of a systemic event defined as a market decline below a threshold \(C\) over a time horizon \(h\) (\citet{acharya2017measuring}). Next, they define multiperiod arithmetic market return between period \(t+1\) and \(t+h\) as \(R_{mt+1:t+h}\) and the systemic event as \({R_{mt+1:t+h}}\). Additionally, they set the horizon \(h\) to 1 month (approx. 22 periods) and the threshold \(C\) to -10\%. Then, the SRISK is defined as the expected capital shortfall conditional on a systemic event:

\begin{equation}
\label{eq:SRISK-1}
\mathrm{SRISK}_{it} = E_{t}(CS_{it+h}|R_{mt+1:t+h}<C) = kE_{t}(D_{it+h}|R_{mt+1:t+h}<C)-(1-k)E_{t}(W_{it+h}|R_{mt+1:t+h}<C)
\end{equation}

Additionally, they assume that in the case of a systemic event debt cannot be renegotiated, what implies that \(E_{t}(D{it+h})|R_{mt+1:t+h}<0=D_{it}\), and finally provide SRISK definition using this assumption:

\begin{equation}
\label{eq:SRISK-2}
\mathrm{SRISK_{it}} = kD_{it}-(1-k)W_{it}(1-\mathrm{LRMES}_{it}) = W_{it}[k\mathrm{LVG}_{it}+(1+k)\mathrm{LRMES}_{it}-1]
\end{equation}
where \(\mathrm{LVG}_{it}\) is the quasi-leverage ratio \((D_{it}+W_{it})/W_{it}\) and \(\mathrm{LRMES}_{it}\) is Long Run Marginal Expected Shortfall, i.e.~Long Run MES, defined as:

\begin{equation}
\label{eq:LRMESit}
\mathrm{LRMES}_{it} = -E_{t}(R_{it+1:t+h}|R_{mt+1:t+h}<C)
\end{equation}
where \(R_{it+1:t+h}\) is the multiperiod arithmetic firm equity return between period \(t+1\) and \(t+h\).

Authors claim that SRISK depend on the size of the firm, its degree of leverage, and its expected equity devaluation conditional on a market decline. SRISK increases when these variables increase. Finally, after pointing that SRISK measure of Equation \ref{eq:SRISK-2} provide a point prediction of the level of capital shortfall a financial entity would experience in case of a systemic
event, they provide the formula for \(\mathrm{SRISK}_{t}\) measure across all firms to construct system-wide measure of financial distress:

\begin{equation}
\label{eq:SRISKt}
\mathrm{SRISK}_{t} = \sum_{i=1}^{N}(\mathrm{SRISK}_{it})_{+}
\end{equation}
where \((x)_{+}\) denotes \(\max(x, 0)\).

At the end, they add that aggregate \(\mathrm{SRISK}_{t}\) should be thought of as the total amount of capital that the government would have to provide to bail out the financial system conditional on the systemic event. At the same time, they admit that they ignore the contribution of negative capital shortfalls (that is capital surpluses) in the computation of aggregate \(\mathrm{SRISK}\).

The last, and probably the most time consuming step to calculate \(\mathrm{SRISK}_{t}\), requires specifying a model for the market and firm returns that can be used to obtain estimators of the LRMES. Nevertheless, a number of different specifications and estimation techniques can be used to obtain this prediction. To construct LRMES predictions, the GARCH-DCC model (\citet{engle2002dynamic}) is used.

\hypertarget{cleveland-feds-systemic-risk-indicator}{%
\subsubsection{Cleveland Fed's Systemic Risk Indicator}\label{cleveland-feds-systemic-risk-indicator}}

\label{sec:Cleveland}

The Cleveland FED's indicator captures the risk of widespread stress in the US banking system (\citet{saldias2013systemic}). This method of computing the systemic risk indicator (cfSRI) is based on the difference, or spread, between two measures of insolvency risk. The first one is an average of default risk across individual banking institutions (average distance-to-default), while the other one is a measure of risk for a weighted portfolio of the same institutions (portfolio distance-to-default).

\begin{equation}
\label{eq:cfSRI}
\mathrm{cfSRI}_{t} = \mathrm{ADD}_{t}-\mathrm{PDD}_{t}
\end{equation}
where \(\mathrm{ADD}_{t}\) is the average distance-to-default, while \(\mathrm{PDD}_{t}\) is the portfolio distance-to-default.

The narrowing of the spread, resulting from the rising insolvency risk of the banking system as a whole, reflects market perceptions of imminent systematic disruption of the banking system. Fragility in the banking system is indicated when falling PDD converges toward ADD (the narrowing of the spread), even when both PDD and ADD are well in positive territory. A spread that is lower than 0.1 for more than two days indicates major financial stress when the average insolvency risk is rising and major banks are stressed by a common factor. When it stays below 0.5 for an extended period of time, it indicates that the markets are signaling major stress about the banking system.

To gauge the level of systemic risk in the banking system, the component parts of cfSRI have to be calculated, i.e.~the average distance-to-default, the portfolio distance-to-default, and then the spread between these two should be interpreted jointly. The average distance-to-default (ADD) reflects the market's perception of the average risk of insolvency among a sample of approximately 100 US banks\footnote{The constituents of an exchange-traded fund (ETF) that reflects the banking system in the aggregate: State Street Global Advisors’ SPDR S\&P Bank ETF, commonly referred to as “KBE".}:

\begin{equation}
\label{eq:ADDt}
\mathrm{ADD}_{t} = \frac{1}{N}\sum_{i=1}^{N} \mathrm{DD}_{i,t}
\end{equation}
where \(\mathrm{DD}_{i}\) is a distance-to-default (DD) for an individual bank \(T\) periods ahead, which is calculated using the Merton model (\citet{Merton1974OnThePricing}) for equity valuation as a European call option on the bank's assets \(A\) at maturity \(T\).

The portfolio distance-to-default (PDD) is a similar measure that is based on options on a weighted portfolio of the same banks\footnote{For this case, it is calculated based on options on an exchange-traded fund (ETF): State Street Global Advisors’ SPDR S\&P Bank ETF (KBE), instead its constituents like it was in the case of ADD.}. It is calculated using options on an exchange-traded fund that reflects the banking system in the aggregate. A decreasing ADD or decreasing PDD indicates the market's perception of rising average insolvency risk in the banking sector.

\hypertarget{catfin}{%
\subsubsection{CATFIN}\label{catfin}}

\label{sec:CATFIN}

CATFIN measure (\citet{allen2012does}) tries to capture the risk of catastrophic losses in the financial system and the statistical approaches to estimating VaR are used for modeling the losses. There are three methodologies used in the VaR estimation: a) direct estimate of the tail risk based on the extreme value distributions, specifically the generalized Pareto distribution (GPD), b) investigation of the shape of the entire return distribution, while providing flexibility of modeling tail thickness and skewness by applying the skewed generalized error distribution (SGED), and c) estimation of VaR based on the left tail of the actual empirical distribution without any assumptions about the underlying return distribution. The first two approaches are known as the parametric methods, whereas the latter one is considered a non-parametric one. In the CATFIN, VaR is estimated at the 99\% confidence level using all three methodologies. Then the first principal component is extracted from the three measures. The CATFIN measure bases on the notion that banks are special for a country's economy and the excess monthly returns on all financial firms are used for the estimation. The extreme returns are defined as the 10\% left tail of the cross-sectional distribution of excess returns on financial firms.

Final formula for CATFIN combines three VaR measures on a monthly level. Principal component analysis (PCA) is used to extract the common component of catastrophic risk embedded in the three proxies in a parsimonious manner, while suppressing potential measurement error associated with the individual VaR measures. This leads to measure the catastrophic risk in the financial system as of month \(t\), denoted CoVaR, as:

\begin{equation}
\label{eq:Carfin}
\mathrm{CATFIN}_{t} = 0.570\upsilon_{\mathrm{GPD}}^{\mathrm{STD}} + 0.5719\upsilon_{\mathrm{SGED}}^{\mathrm{STD}} + 0.5889\upsilon_{\mathrm{NP}}^{\mathrm{STD}}
\end{equation}
where \(\upsilon_{\mathrm{GPD}}^{\mathrm{STD}}\), \(\upsilon_{\mathrm{SGED}}^{\mathrm{STD}}\), and \(\upsilon_{\mathrm{NP}}^{\mathrm{STD}}\) correspond to the standardized VaR measures based on the GPD, the SGED, and the non-parametric methods, respectively.

\hypertarget{composite-indicator-of-systemic-stress-ciss}{%
\subsubsection{Composite indicator of systemic stress (CISS)}\label{composite-indicator-of-systemic-stress-ciss}}

\label{sec:CISS}

A Composite Indicator of Systemic Stress (CISS) is a broad measure of systemic risk in the financial system. The financial system can be divided into three main building blocks: markets, intermediaries, and infrastructures. Each of these building blocks can be split into specific segments. The financial markets segment can be separated into individual markets like money, equity, bond, currency, and derivatives. Within financial intermediaries, the most important are banks, and insurance companies. The market infrastructure is composed of payment settlement and clearing systems.

CISS aggregates financial stress at two levels: it first computes five segment-specific stress subindices, and then aggregates these five subindices into the final composite stress index. It mostly relies on realized asset return volatilities and on risk spreads to capture the main symptoms of financial stress in the various market segments. The subindices are aggregated analogously to the aggregation of individual asset risks into overall portfolio risk by taking into account the cross-correlations between all individual asset returns and not only their variances. It is essential for the purpose of constructing of the systemic risk indicator to allow for time-variation in the cross-correlation structure between subindices. In this case, the CISS puts more weight on situations in which high stress prevails in several market segments at the same time. The stronger financial stress is correlated across subindices, the more widespread is the state of financial instability according to the ``horizontal view'' of the definition of systemic stress. The second element of the aggregation scheme potentially featuring systemic risk is the fact that the subindex weights can be determined on the basis of their relative importance for real economic activity. This specific feature in the design of the CISS not only offers a way to capture the ``vertical view'' of systemic stress, but in doing so it also implicitly accounts for country differences in the structure of their financial systems as long as these actually matter for the transmission of financial stress to the real economy. For the Euro area the subindex weights are: money market 15\%, bond market 15\%, equity market 25\%, financial intermediaries 30\%, and foreign exchange market 15\%.

\hypertarget{comparison-and-forecasting-ability-of-ivrvsri}{%
\subsection{Comparison and Forecasting ability of IVRVSRI}\label{comparison-and-forecasting-ability-of-ivrvsri}}

\label{sec:ForecastingabilityofIVRVSRI}

\hypertarget{correlation-matrix-and-rolling-correlation}{%
\subsubsection{Correlation matrix and rolling correlation}\label{correlation-matrix-and-rolling-correlation}}

\label{sec:CorrelationMatrixAndRollingCorrelation}

In order to compare existing SRIs with our IVRVSRI, first we visualize their levels, returns and descriptive statistics of weekly returns which are used in regressions in the next step. We also calculate the correlation matrices for four distinct benchmarks SRIs, IVRVSRI and S\&P500: for weekly returns of all of them and between returns of S\&P500 index and lagged weekly returns of SRIs.

\hypertarget{forecasting-ability}{%
\subsubsection{Forecasting ability}\label{forecasting-ability}}

\label{sec:forecasting ability}

\hypertarget{simple-regression-model}{%
\paragraph{Simple Regression Model}\label{simple-regression-model}}

\label{sec:Simple Regression Model}

~\newline

The first model (Equation \ref{eq:SRegM-1}) is be based on simple regression of S\&P 500 index weekly returns regressed on the lagged weekly returns of the given SRI (either a benchmark one, or the IVRVSRI):

\begin{equation}
\label{eq:SRegM-1}
r_{t}^{w,\mathrm{SP500}} = \beta_0 + \sum_{k=1}^{p}\beta_k^{i}r_{t-k}^{w,\mathrm{SRI}_i}+\varepsilon_{t}
\end{equation}
where:

\(r_{t}^{w,\mathrm{SP500}}\) - weekly return of S\&P 500 index on day \(t\),

\(\beta_{k}^{i}\) - sensitivity of \(r_{t}^{w, \mathrm{SP500}}\) to lag \(k\) of the given \(\mathrm{SRI}_i\) return,

\(r_{t-k}^{w,\mathrm{SRI}_i}\) - lag \(k\) of weekly return of the given \(\mathrm{SRI}_{i}\) on day \(t\).

\vspace{3mm}

Additionally, forecasting abilities of each SRIs are investigated jointly:

\begin{equation}
\label{eq:SRegM-2}
r_{t}^{w,\mathrm{SP500}} = \beta_0 + \sum_{k = 1}^{p}\sum_{i = 1}^{N}\beta_{k}^{i}r_{t-k}^{w, \mathrm{SRI}_{i}}+\varepsilon_{t}
\end{equation}
where:

\(\beta_{k}^{i}\) - sensitivity of \(r_{t}^{w, \mathrm{SP500}}\) to lag \(k\) of the given \(\mathrm{SRI}_i\) return,

\(r_{t-k}^{w,\mathrm{SRI}_i}\) - lag \(k\) of weekly return of the given \(\mathrm{SRI}_{i}\) on day \(t\).

\hypertarget{quasi-quantile-regression-model}{%
\paragraph{Quasi-quantile Regression Model}\label{quasi-quantile-regression-model}}

\label{sec:Quasi Quantile  Regression Model}

~\newline

Quasi-quantile Regression Model is our innovative idea where we estimate models described with Equation \ref{eq:SRegM-1} and Equation \ref{eq:SRegM-2} only for such values \(r_{t}^{w,\mathrm{SP500}}\) which fulfill the following conditions:

\begin{itemize}
\item
  \(r_{t}^{w,\mathrm{SP500}}\) \textless{} \(\bar{r}_{\mathrm{SP500},t}^{w}\)
\item
  \(r_{t}^{w,\mathrm{SP500}}\) \textless{} 1st quartile of \(r_{t}^{w,\mathrm{SP500}}\)
\item
  \(r_{t}^{w,\mathrm{SP500}}\) \textless{} 1st decile of \(r_{t}^{w,\mathrm{SP500}}\)
\item
  \(r_{t}^{w,\mathrm{SP500}}\) \textless{} 5th percentile of \(r_{t}^{w,\mathrm{SP500}}\)
\item
  \(r_{t}^{w,\mathrm{SP500}}\) \textless{} 2.5th percentile of \(r_{t}^{w,\mathrm{SP500}}\)
\item
  \(r_{t}^{w,\mathrm{SP500}}\) \textless{} 1st percentile of \(r_{t}^{w,\mathrm{SP500}}\)
\end{itemize}

Hence, model specification for this approach is given by:

\begin{equation}
\label{eq:qQRegM-1}
r_{t}^{w,\mathrm{SP500}}(p) = \beta_0 + \sum_{k = 1}^{p}\beta_{k}^{i}r_{t-k}^{w, \mathrm{SRI}_{i}}+\varepsilon_{t}
\end{equation}
where \(r_{t}^{w,\mathrm{SP500}}(p)\) is weekly return of S\&P 500 index on day \(t\) conditional on percentile \(p\) of its distribution.

\hypertarget{quantile-regression-model}{%
\paragraph{Quantile Regression Model}\label{quantile-regression-model}}

\label{sec:QuantileRegressionModel}

~\newline

Following \citet{caporin2021traffic}, we estimate the quantile regression model (Equation \ref{eq:QRegM-1}) in order to present a comprehensive picture of the forecasting ability of aggregated systemic risk indicators on the equity market conditioned on the location of the equity index return over its density.

\begin{equation}
\label{eq:QRegM-1}
Q_{\tau}(r_{t}^{w,\mathrm{SP500}}) = \beta_0(\tau) + \sum_{k = 1}^{p}\beta_{k}(\tau)r_{t-k}^{w, \mathrm{SRI}_i}+\varepsilon_{t}
\end{equation}
where:

\(Q_{\tau}(r_{t}^{w,\mathrm{SP500}})\) - \(\tau\) quantile of S\&P 500 returns on day \(t\),

\(r_{t-k}^{w,\mathrm{SRI}_i}\) - lag \(k\) of return of the given \(\mathrm{SRI}_i\),

\(\beta_{k}(\tau)\) - percentage change in \(\tau\) quantile of the S\&P500 index returns produced by the change in the predictor \(k\) intervals earlier.

\hypertarget{data}{%
\section{Data}\label{data}}

\label{sec:Data}

\begin{figure}[H]

\includegraphics[width=1\linewidth]{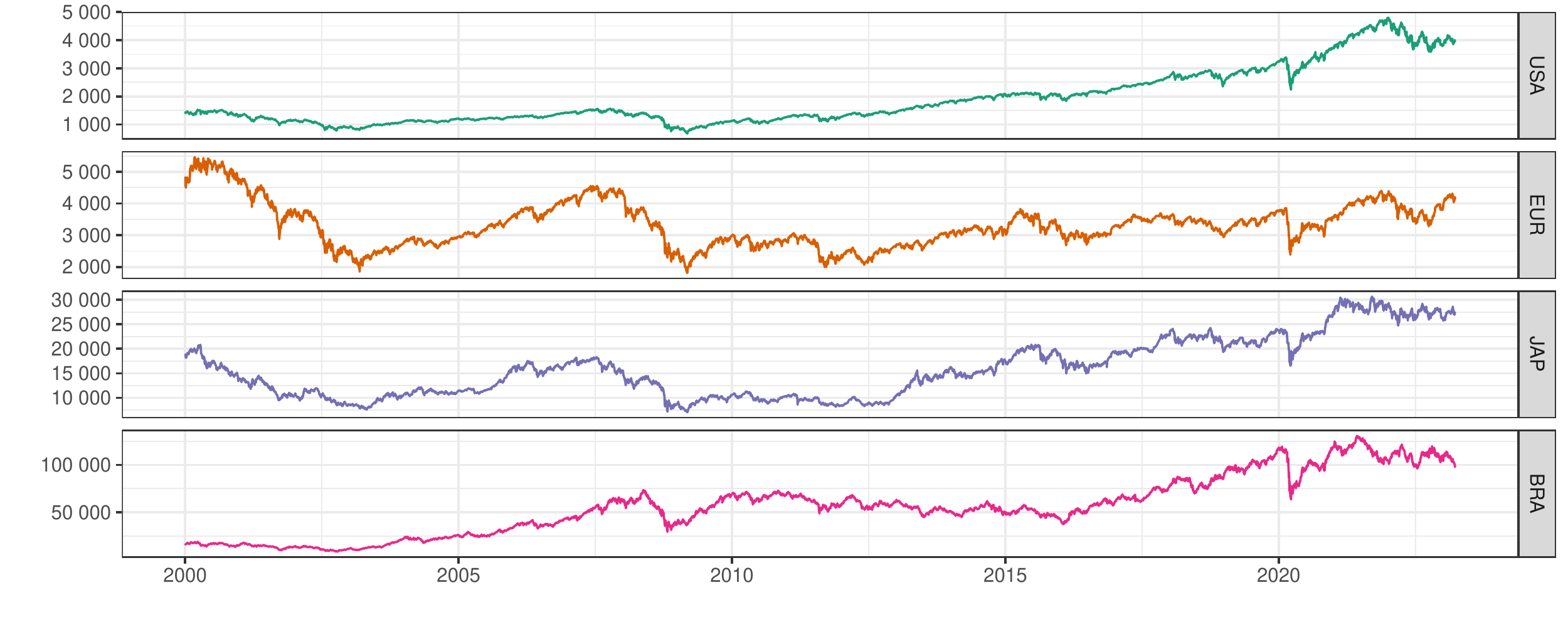}

\caption{The fluctuations of S\&P500, EuroStoxx50, Nikkei225 and Bovespa indices between 2000 and 2023. \label{fig:EquityLines}}
\scriptsize \textrm{Note: The main equity indices for USA, Europe, Japan and Brazil in the period between 2000 and 2023.}
\end{figure}

\begin{figure}[H]

\includegraphics[width=1\linewidth]{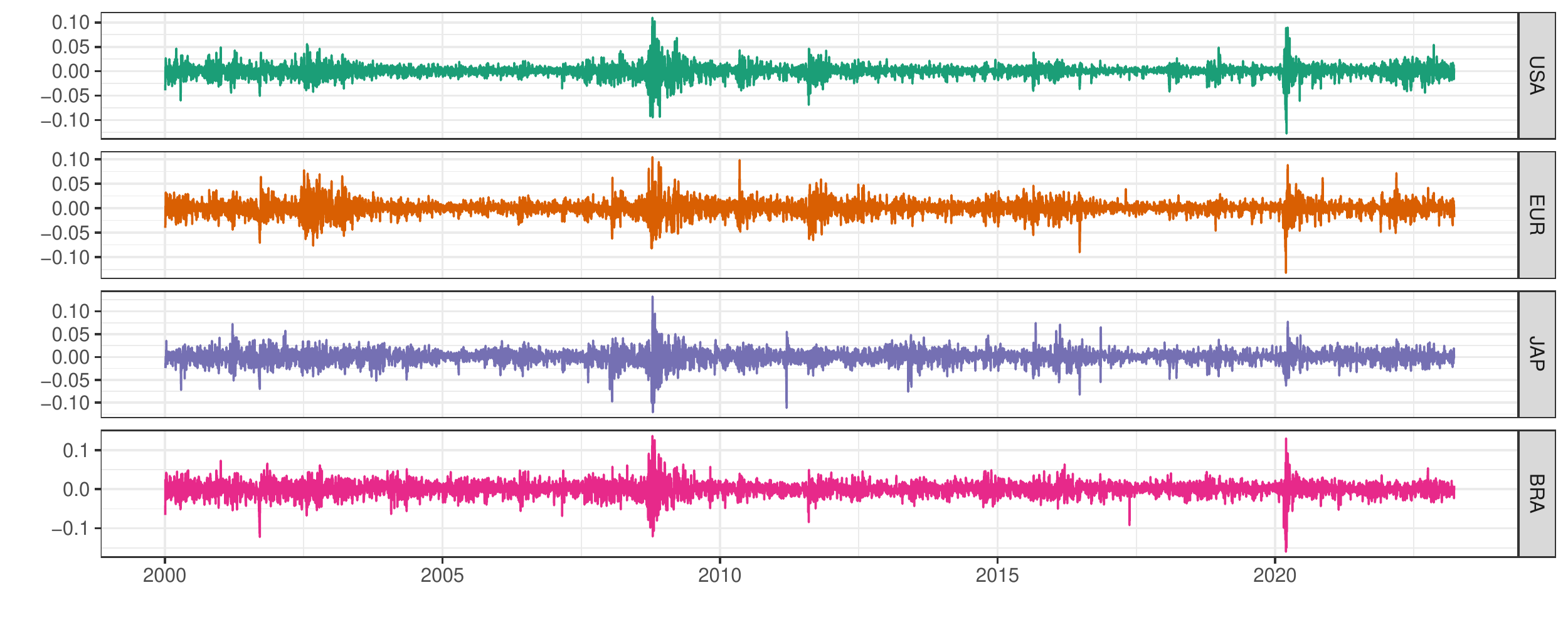}

\caption{Returns of S\&P500, EuroStoxx50, Nikkei225 and Bovespa indices between 2000 and 2023. \label{fig:returns}}
\scriptsize \textrm{Note: Returns of the main equity indices for USA, Europe, Japan and Brazil in the period between 2000 and 2023.}
\end{figure}

Our data set is based on daily data for volatility indices (VIX, VSTOXX, VNKY, and VXEWZ) and daily price and market cap data for equity indices (S\&P500, EuroStoxx50, Nikkei 225, Bovespa) in the period between 2000 to 2023. Figure \ref{fig:EquityLines} presents the fluctuations of the analyzed times series, while Figure \ref{fig:returns} fluctuations of returns. Figure \ref{fig:EquityLines} informs us about different magnitude of upward and downward movements on analyzed markets, while Figure \ref{fig:returns} additionally visualize volatility clustering with high and low volatility periods indicating calm and more stressfull periods of time.

\begin{figure}[H]

\includegraphics[width=1\linewidth]{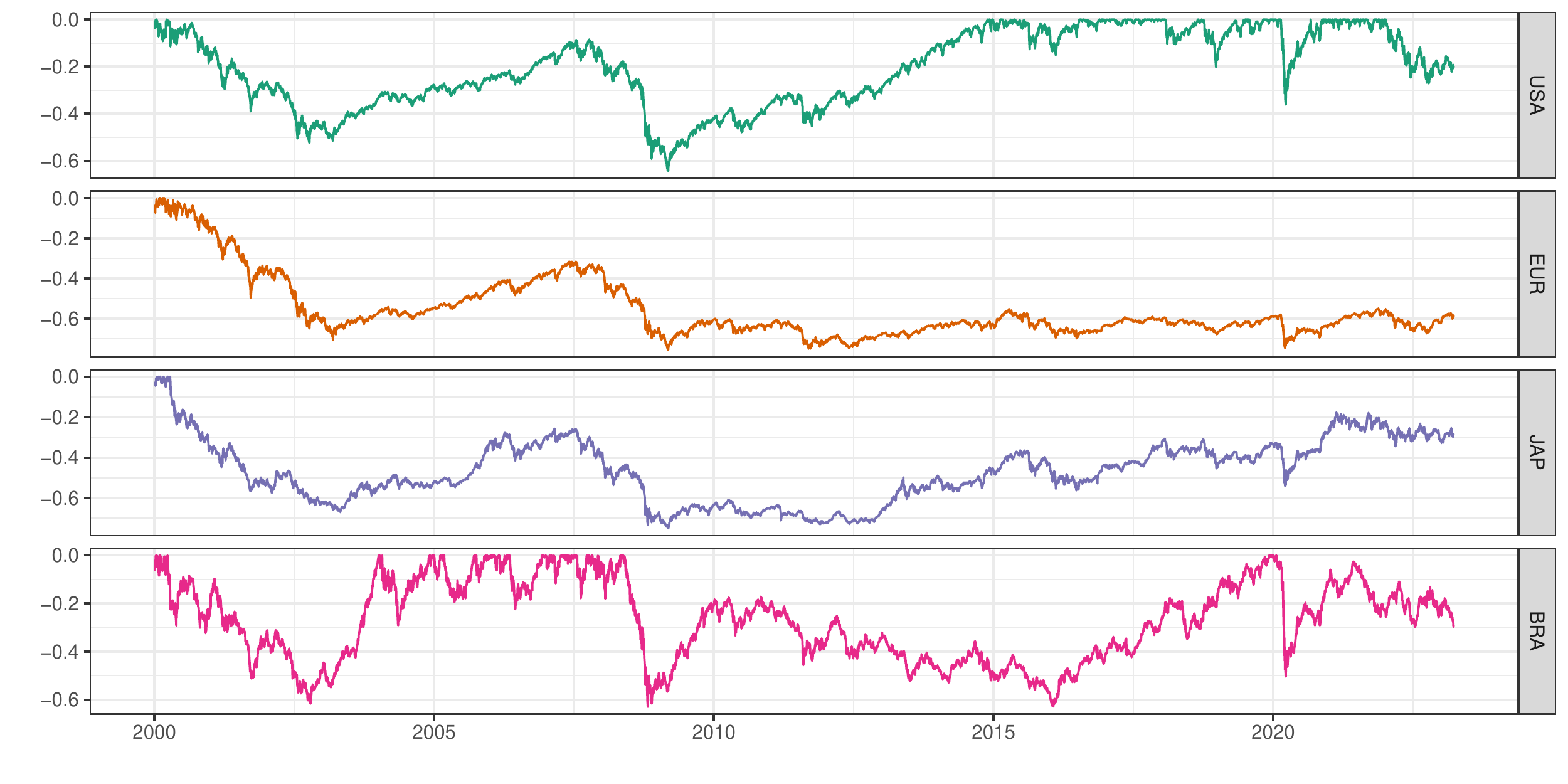}

\caption{Drawdowns of S\&P500, EuroStoxx50, Nikkei225 and Bovespa indices between 2000 and 2023. \label{fig:drawdowns}}
\scriptsize \textrm{Note: Panel (1) presents drawdowns for S\&P500 index prices. Panel (2) presents drawdowns for EuroStoxx50 index prices. Panel (3) presents drawdowns for Nikkei225 index prices. Panel (4) presents drawdowns for iShares Brazil ETF (EWZ) index prices.}
\end{figure}

Drawdowns of analyzed equity indices, depicted on Figure \ref{fig:drawdowns}, show the length of the most important turmoils and additionally visualize their speed and magnitude.

Descriptive statistics of returns, presented in Table \ref{tab:desc-stats}, confirm the well-known fact about equity returns, i.e.~high kurtosis, negative skewness and associated non-normality of returns.

\begin{table}[!h]

\caption{\label{tab:desc-stats}Descriptive statistics of equity indices returns.}
\centering
\fontsize{8}{10}\selectfont
\begin{threeparttable}
\begin{tabular}[t]{>{\raggedright\arraybackslash}p{11em}rrrr}
\toprule
statistic & USA & EUR & JAP & BRA\\
\midrule
\textbf{nobs} & 5860 & 5860 & 5860 & 5860\\
\textbf{NAs} & 1 & 1 & 2 & 1\\
\textbf{Minimum} & -0.13 & -0.13 & -0.12 & -0.16\\
\textbf{1. Quartile} & 0 & -0.01 & -0.01 & -0.01\\
\textbf{Mean} & 0 & 0 & 0 & 0\\
\textbf{Median} & 0 & 0 & 0 & 0\\
\textbf{3. Quartile} & 0.01 & 0.01 & 0.01 & 0.01\\
\textbf{Maximum} & 0.11 & 0.1 & 0.13 & 0.14\\
\textbf{Stdev} & 0.01 & 0.01 & 0.01 & 0.02\\
\textbf{Skewness} & -0.38 & -0.19 & -0.4 & -0.39\\
\textbf{Kurtosis} & 10.17 & 5.97 & 6.82 & 7.08\\
\textbf{Norm.} & 0 & 0 & 0 & 0\\
\bottomrule
\end{tabular}
\begin{tablenotes}[para]
\small
\item \footnotesize{Note: 'Norm.' denotes p-value of the Jareque-Bera test for normality.}
\end{tablenotes}
\end{threeparttable}
\end{table}

In order to calculate the proper weights in IVSRI, RVSRI and IVRVSRI indicators, we decided to use market capitalization data for each of the equity indices used (Table \ref{tab:MarketCap}).

\begin{table}[H]

\caption{\label{tab:MarketCap}Market capitalisation for S\&P500, EuroStoxx50, Nikkei225 and Bovespa indices.}
\centering
\fontsize{8}{10}\selectfont
\begin{threeparttable}
\begin{tabular}[t]{>{\raggedright\arraybackslash}p{11em}>{\raggedleft\arraybackslash}p{11em}>{\raggedleft\arraybackslash}p{11em}}
\toprule
Indices & MarketCap & Weights\\
\midrule
\textbf{S\&P500} & 35.6T & 77.7\%\\
\textbf{EuroStoxx50} & 3.7T & 8.1\%\\
\textbf{Nikkei225} & 5.5T & 12\%\\
\textbf{Bovespa} & 1T & 2.2\%\\
\textbf{TOTAL} & 45.9T & 100\%\\
\bottomrule
\end{tabular}
\begin{tablenotes}
\small
\item \footnotesize{Note: Market Capitalization for equity indices were downloaded on 2021-06-21 from:} 
\item \footnotesize{S\&P500 index: \url{https://ycharts.com/indicators/sp\_500\_market\_cap}}
\item \footnotesize{EuroStoxx50: \url{https://en.wikipedia.org/wiki/EURO\_STOXX\_50}}
\item \footnotesize{Nikkei 225: \url{https://www.bloomberg.com/quote/NKY:IND}}
\item \footnotesize{Bovespa: \url{https://en.wikipedia.org/wiki/B3\_(stock\_exchange)}}
\end{tablenotes}
\end{threeparttable}
\end{table}

\hypertarget{results}{%
\section{Results}\label{results}}

\label{Results}

Based on the logic ``keep it simple'', we want to check if it is possible to create Systemic Risk Indicator based on widely available (most often publicly available and free of charge) volatility risk measures which can have similar properties as systemic risk indicators introduced in highly cited papers (\citet{brownlees2017srisk}, \citet{acharya2017measuring}, \citet{romer2017new} or \citet{engle2018much}) or in the most recent study of \citet{caporin2021traffic}. In the Results section, we present Figures and map charts visualizing systemic risk indicators and theirs components.

\hypertarget{ivrvsris-on-the-country-level}{%
\subsection{IVRVSRIs on the country level}\label{ivrvsris-on-the-country-level}}

\label{IVRVSRIsOnTheCountryLevel}

\begin{figure}[H]

\includegraphics[width=1\linewidth]{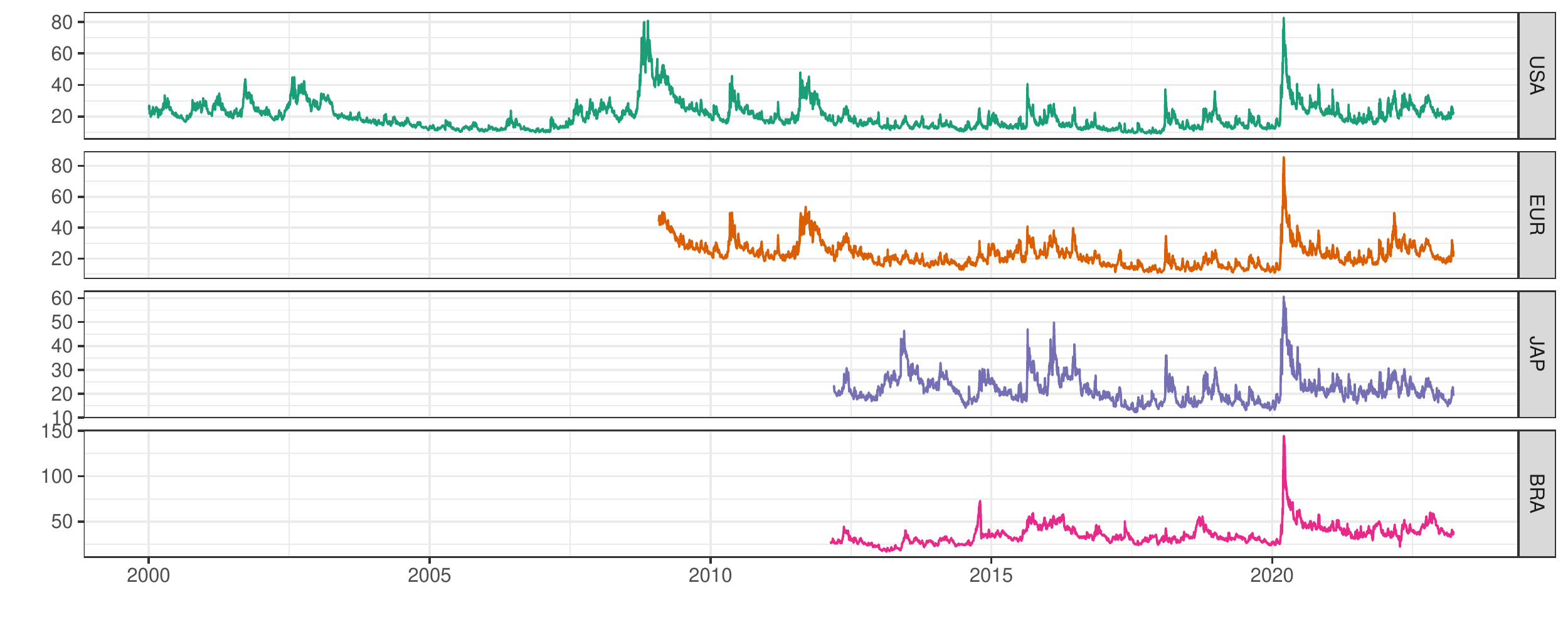}

\caption{Implied Volatility indices for S\&P500, EuroStoxx50, Nikkei225 and Bovespa between 2000 and 2023 \label{fig:IVindices}}
\scriptsize Note: Panel (1) presents VIX index calculated based on S\&P500 index options series. Panel (2) presents VStoxx index calculated based on EuroStoxx50 index options series. Panel (3) presents VNKY index calculated based on Nikkei225 index options series. Panel (4) presents VXEWZ index calculated based on the iShares Brazil ETF (EWZ) index options series. 
\end{figure}

Figure \ref{fig:IVindices} shows the fluctuations of IV indices for each country separately and shows the most significant turmoils affecting the equity market in each country under investigation, i.e.~GFC (20087-2009), COVID pandemic (March 2020), and a few of lower magnitude like Eurozone debt crisis (2009-2014), and turmoils in August 2015, February 2018 and November-December 2018.

\begin{figure}[H]

\includegraphics[width=1\linewidth]{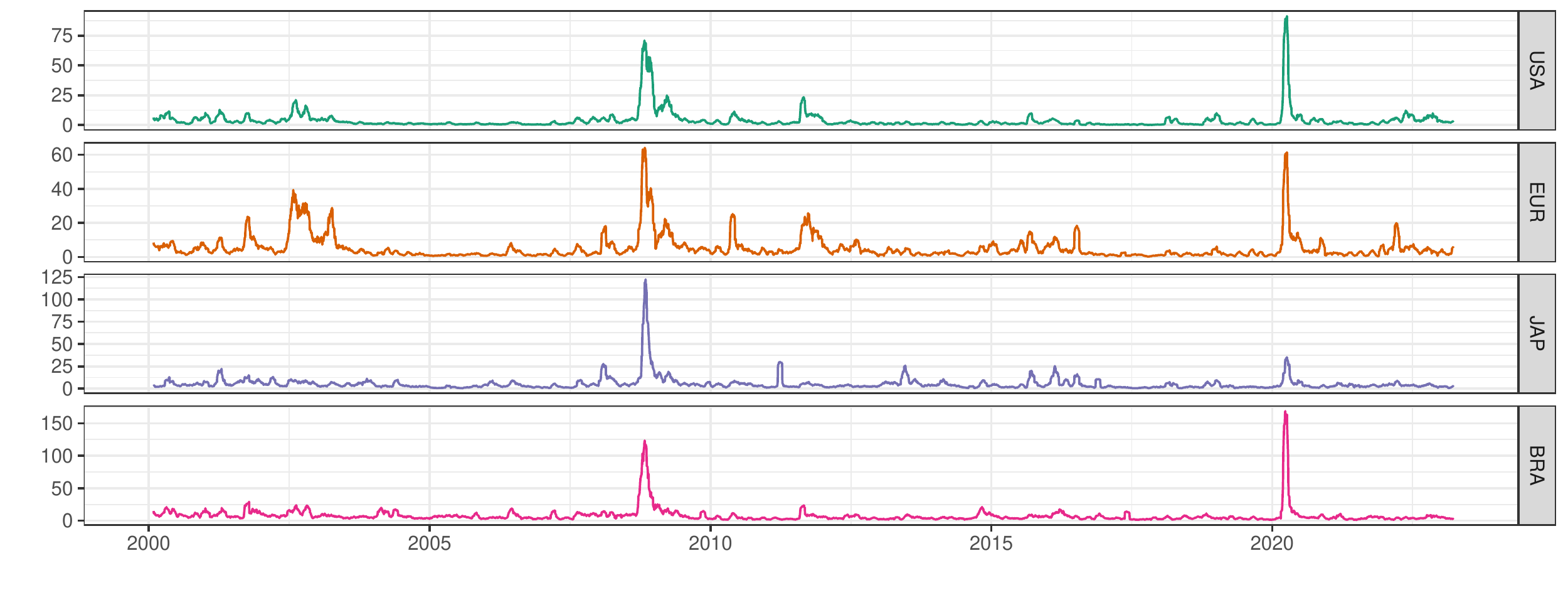}

\caption{Realized Volatility indices for S\&P500, EuroStoxx50, Nikkei225 and Bovespa between 2000 and 2023 \label{fig:RVindices}}
\scriptsize Note: Panel (1) presents RV index calculated based on S\&P500 index prices based on the formula \ref{eq:RVi}. Panel (2) presents VStoxx index calculated based on EuroStoxx50 index prices based on the formula \ref{eq:RVi}. Panel (3) presents VNKY index calculated based on Nikkei225 index prices based on the formula \ref{eq:RVi}. Panel (4) presents VXEWZ index calculated based on the iShares Brazil ETF (EWZ) index prices based on the formula \ref{eq:RVi}.
\end{figure}

On the other hand, Figure \ref{fig:RVindices} presents RV indices for each country separately. Comparing Figure (\ref{fig:IVindices} with Figure \ref{fig:RVindices}) shows that the anticipated reaction (IV indices in Figure \ref{fig:IVindices}) to the current market stress is not always the same as the current reaction revealed in realized volatility of returns (IV versus RV for Japan during Covid pandemic in March 2020).

\begin{figure}[H]

\includegraphics[width=1\linewidth]{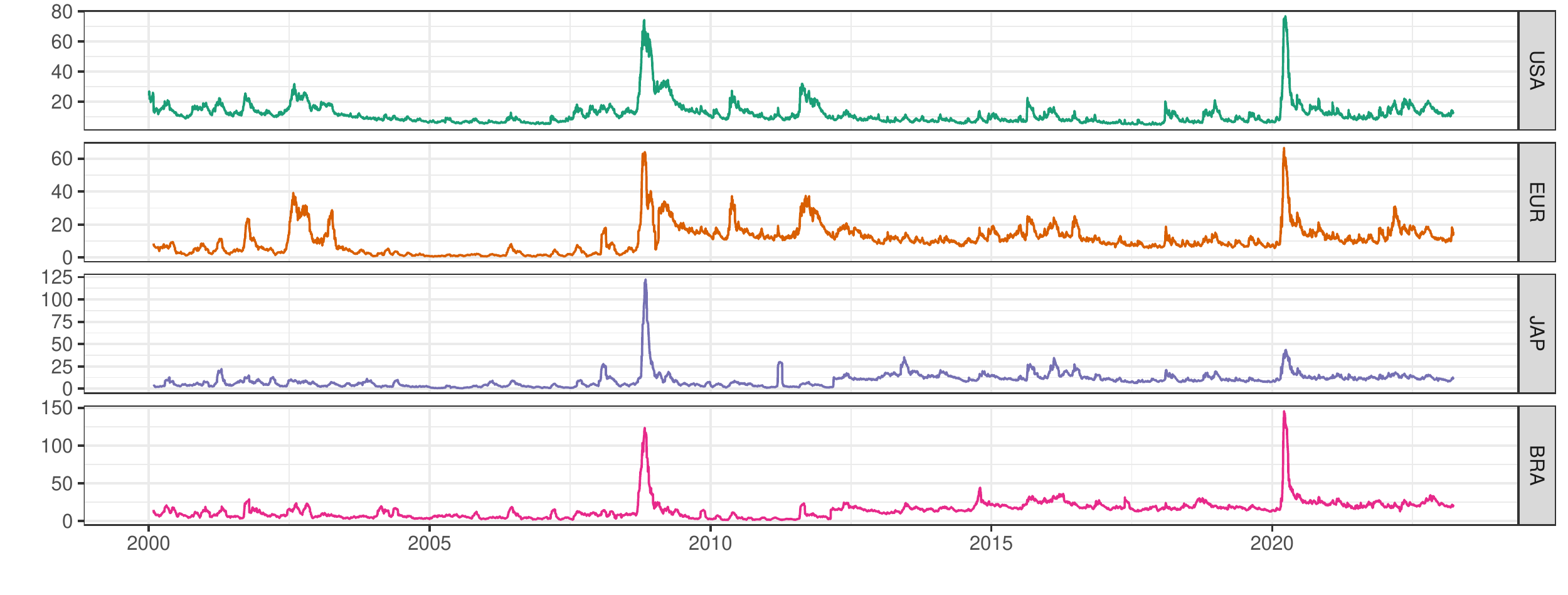}

\caption{IVRVSRI on the country level separately for S\&P500, EuroStoxx50, Nikkei225 and Bovespa between 2000 and 2023\label{fig:IVRVSRIindices}}
\scriptsize Note: Panel (1) presents IVRVSRI for the USA calculated based on VIX index and S\&P500 index prices based on the formula \ref{eq:IVRVSRI_i}. Panel (2) presents IVRVSRI for Eurozone calculated based on the VSTOXX index and EuroStoxx50 index prices based on the formula \ref{eq:IVRVSRI_i}. Panel (3) presents IVRVSRI for Japan calculated based on the VNKY index and Nikkei225 index prices based on the formula \ref{eq:IVRVSRI_i}. Panel (4) presents  IVRVSRI calculated based on the VXEWZ index and the iShares Brazil ETF (EWZ) index prices based on the formula \ref{eq:IVRVSRI_i}.
\end{figure}

Overall, our results show that the magnitude of reactions to the risk events varies across countries. Analyzing IVRVSRI indications on the country levels presented on Figure \ref{fig:IVRVSRIindices}, we observe a very weak reaction of Japanese markets to COVID-19 pandemic in March 2020 in comparison to the USD and Eurozone, and literally no reaction of Japanese and Brazilian markets to the European sovereign debt crisis in 2009-2014. Only in the case of the GFC 2007-2009 all analyzed markets reacted strongly but the persistence of the crisis was not the same (Figure \ref{fig:IVRVSRIindices}). Brazil and Japan recovered quickly with regard to the speed of the decrease of IVRVSRI indications while the USA and Europe were struggling much longer.

\begin{figure}[H]

\includegraphics[width=1\linewidth]{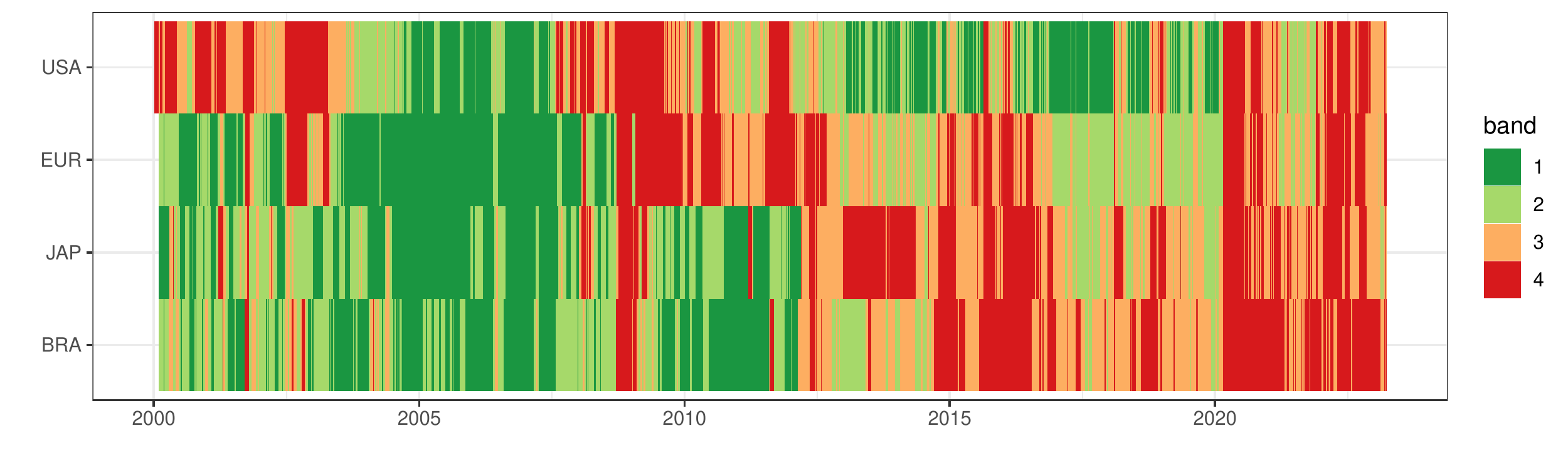}

\caption{Quartile colour based IVRVSRI map chart on the country level \label{fig:MapIVRVSRIcountry}}
\scriptsize Note: Panel (1) presents colored map chart indicating quartiles of $\mathrm{IVRVSRI}_i$ for the USA calculated based on VIX index and S\&P500 index prices based on the formula \ref{eq:IVRVSRI_i}. Panel (2) presents colored map chart indicating quartiles of $\mathrm{IVRVSRI}_i$ for Eurozone calculated based on the VSTOXX index and EuroStoxx50 index prices based on the formula \ref{eq:IVRVSRI_i}. Panel (3) presents colored map chart indicating quartiles of $\mathrm{IVRVSRI}_i$ for Japan calculated based on the WNKY index and Nikkei225 index prices based on the formula \ref{eq:IVRVSRI_i}. Panel (4) presents colored map chart indicating quartiles of $\mathrm{IVRVSRI}_i$ calculated based on the VXEWZ index and the iShares Brazil ETF (EWZ) index prices based on the formula \ref{eq:IVRVSRI_i}. Quartiles on map chart are indicated with green-red scale, where green indicates the 1st quartile (the lowest one) while red colour indicates the 4th quartile (the highest one).
\end{figure}

Next, Figure \ref{fig:MapIVRVSRIcountry} shows the map chart with colored quartile levels of \(\mathrm{IVRVSRI}\) indications on the country level. It shows that in the case of Eurozone, the GFC extended into the debt crisis and lasted with a small break in 2014 until 2016. In general, before the GFC the Eurozone, Japanese and Brazilian markets were more resilient than the American one to worldwide turmoils while the situation reversed after the Eurozone sovereign debt crisis, with Brazil and Japan being the least resilient in that period among all analyzed countries.

\hypertarget{ivrvsris-on-the-global-level}{%
\subsection{IVRVSRIs on the global level}\label{ivrvsris-on-the-global-level}}

\label{IVRVSRIsOnTheGlobalLevel}

\begin{figure}[H]

\includegraphics[width=1\linewidth]{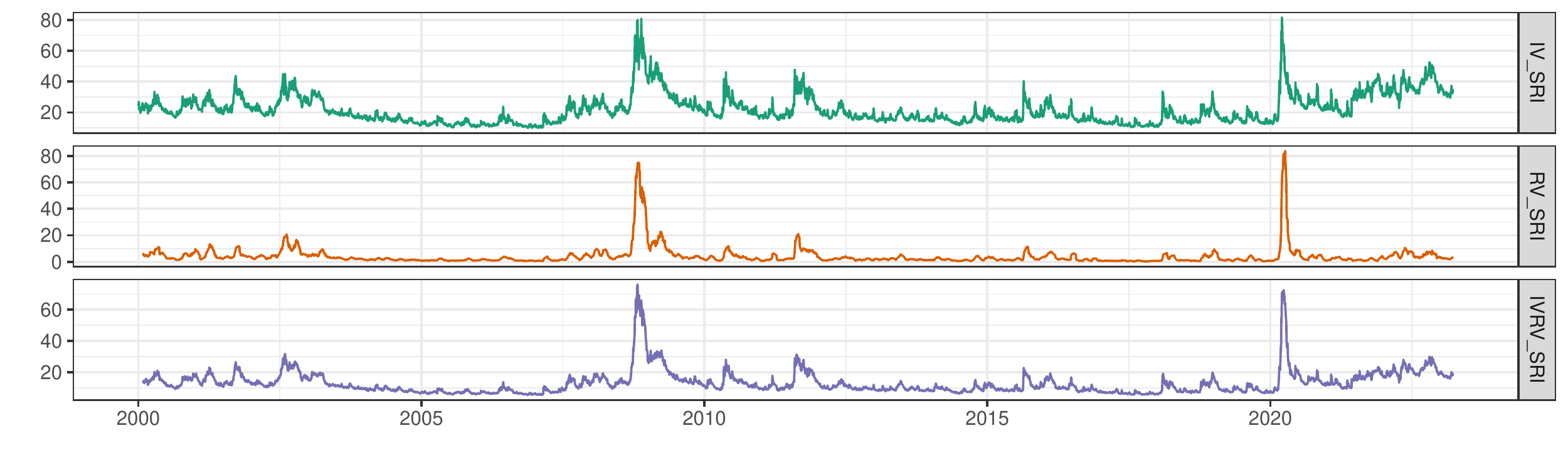}

\caption{IVSRI RVSRI and IVRVSRI on the global level\label{fig:IV-RV-IVRV-global}}
\scriptsize Note: Panel (1) presents $\mathrm{IVSRI}$ on the global level calculated based on  on the formula \ref{eq:IVSRI}. Panel (2) presents $\mathrm{RVSRI}$ on the global level calculated based on the formula \ref{eq:RVSRI}. Panel (3) presents $\mathrm{IVRVSRI}$ on the global level calculated based on the formula \ref{eq:IVRVSRI}.
\end{figure}

Figure \ref{fig:IV-RV-IVRV-global} shows the aggregated results for \(\mathrm{IVRVSRI}\) and its components (\(\mathrm{IVSRI}\) and \(\mathrm{RVSRI}\)) on the global level. We can see that after aggregation of the country specific indices all the major financial crises are indicated and additionally we can observed their severity. GFC and Covid were the most severe turmoils, but other ones line the end of downward trend after the Dotcom bubble (2002-2003) and Eurozone debt crisis (2009-2014) are revealed as well. What is more, the reaction of IVSRI and RVSRI components on the global level to the above mentioned turmoils differs with regard to the magnitude of their reaction. Most often, the fear revealed in IVSRI (Panel (1) of Figure \ref{fig:IV-RV-IVRV-global}), especially in case of less severe turmoils (Eurozone debt crisis or the bottom of the Dotcom bubble), was not realized in the same magnitude of RVSRI indications (Panel (2) of Figure \ref{fig:IV-RV-IVRV-global}).

\begin{figure}[H]

\includegraphics[width=1\linewidth]{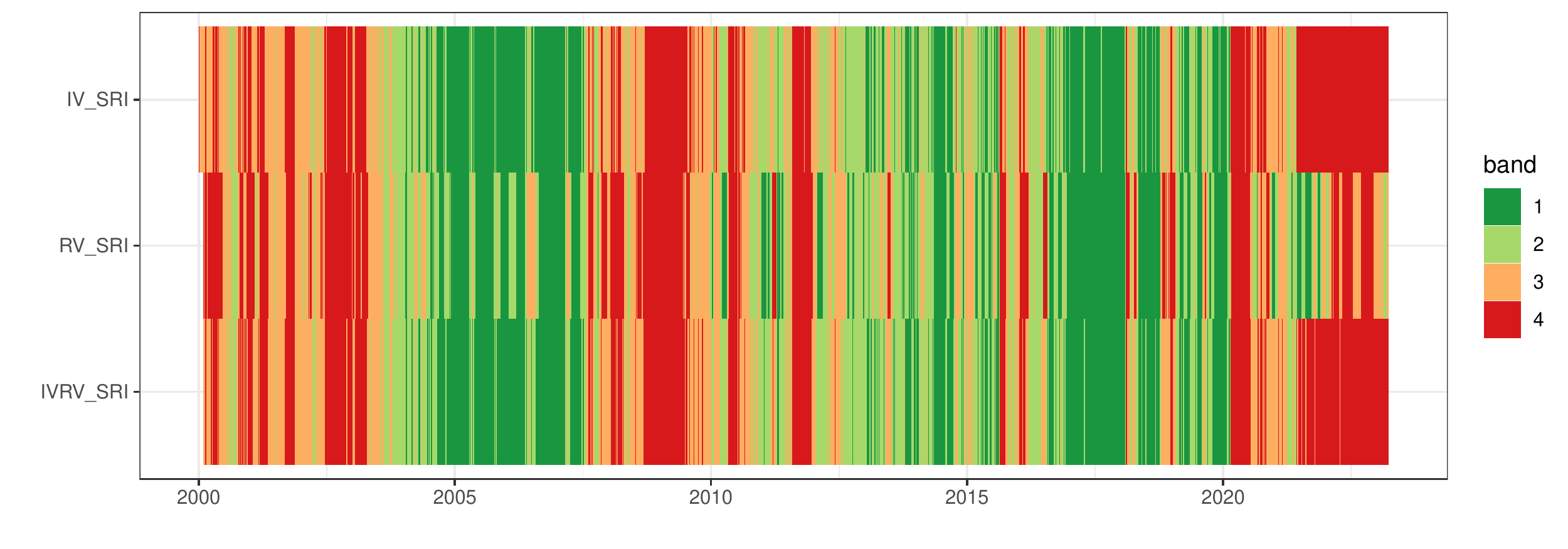}

\caption{Quartile colour based IVSRI, RVSRI and IVRVSRI map on the global level\label{fig:MapIV-RV-IVRV-global}}
\scriptsize Note: Panel (1) presents colored map chart indicating quartiles of $\mathrm{IVSRI}_i$ on the global level calculated based on the formula \ref{eq:IVSRI}. Panel (2) presents colored map chart indicating quartiles of $\mathrm{RVSRI}_i$ on the global level calculated based on the formula \ref{eq:RVSRI}. Panel (3) presents colored map chart indicating quartiles of $\mathrm{IVRVSRI}_i$ on the global level calculated based on the formula \ref{eq:IVRVSRI}. Quartiles on map chart are indicated with green-red scale , where green indicates the 1st quartile (the lowest one) while red colour indicates the 4th quartiel (the highest one).
\end{figure}

Figure \ref{fig:MapIV-RV-IVRV-global} presents a colored map chart indicating quartiles of \(\mathrm{IVSRI}\), \(\mathrm{RVSRI}\), and \(\mathrm{IVRVSRI}\) on the global level stressing the major turmoils on the aggregated level.

The IVSRI and RVSRI show slightly different risk levels in the ``transition'' periods when systemic risk changes. In general, we can state that the reaction of the implied-volatility-based metrics is faster than the realized volatility one, which is something we have expected. Moreover, the correlation between the IV-based indicator and the general systemic risk indicator (\(\mathrm{IVRVSRI}\)) is higher than that of the RV-based ones. At the same time, the general systemic risk indicator (\(\mathrm{IVRVSRI}\)) is a better indicator of systemic risk than any individual indicator based on only one measure of volatility (\(\mathrm{RVSRI}\) or \(\mathrm{IVSRI}\)), and this result is robust even after the change of the weights of the \(\mathrm{RVSRI}\) and \(\mathrm{IVSRI}\) in the general systemic risk measure.

\begin{figure}[H]

\includegraphics[width=1\linewidth]{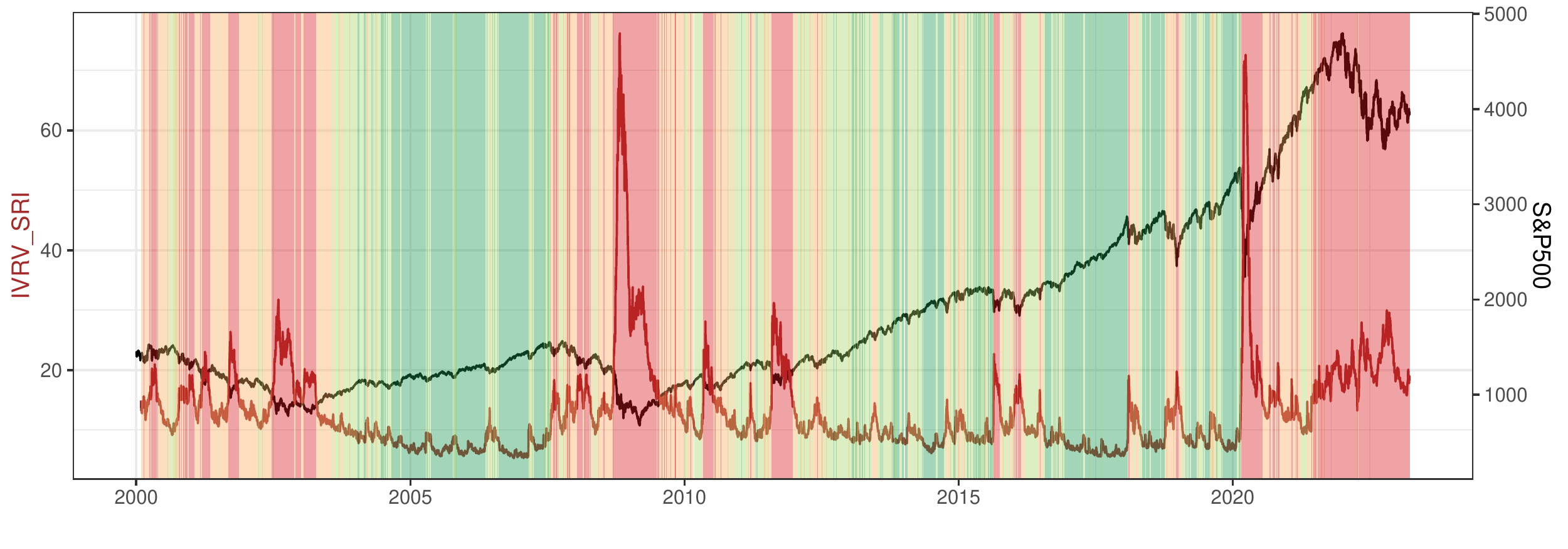}

\caption{IVRVSRI and S\&P500 index \label{fig:SP500-IVRVSRI-map2} on colored map chart with quartiles of IVRVSRI.}
\scriptsize Note: The flucuations of S\&P500 index shows and IVRVSRI on the global level on the background of colored map chart with quartiles of IVRVSRI.
\end{figure}

Figure \ref{fig:SP500-IVRVSRI-map2} depicts the comparison of fluctuations of S\&P500 index and IVRVSRI on the global level. It clearly shows that each major financial turmoil was reflected on our IVRVSRI almost immediately informing market participants about increased level of stress.

\newpage

\hypertarget{comparison-between-sris-and-sp500-index}{%
\subsection{Comparison between SRIs and S\&P500 index}\label{comparison-between-sris-and-sp500-index}}

\label{ComparisonBetweenSRIsAndS&P500Index}

\begin{figure}[H]

\includegraphics[width=1\linewidth]{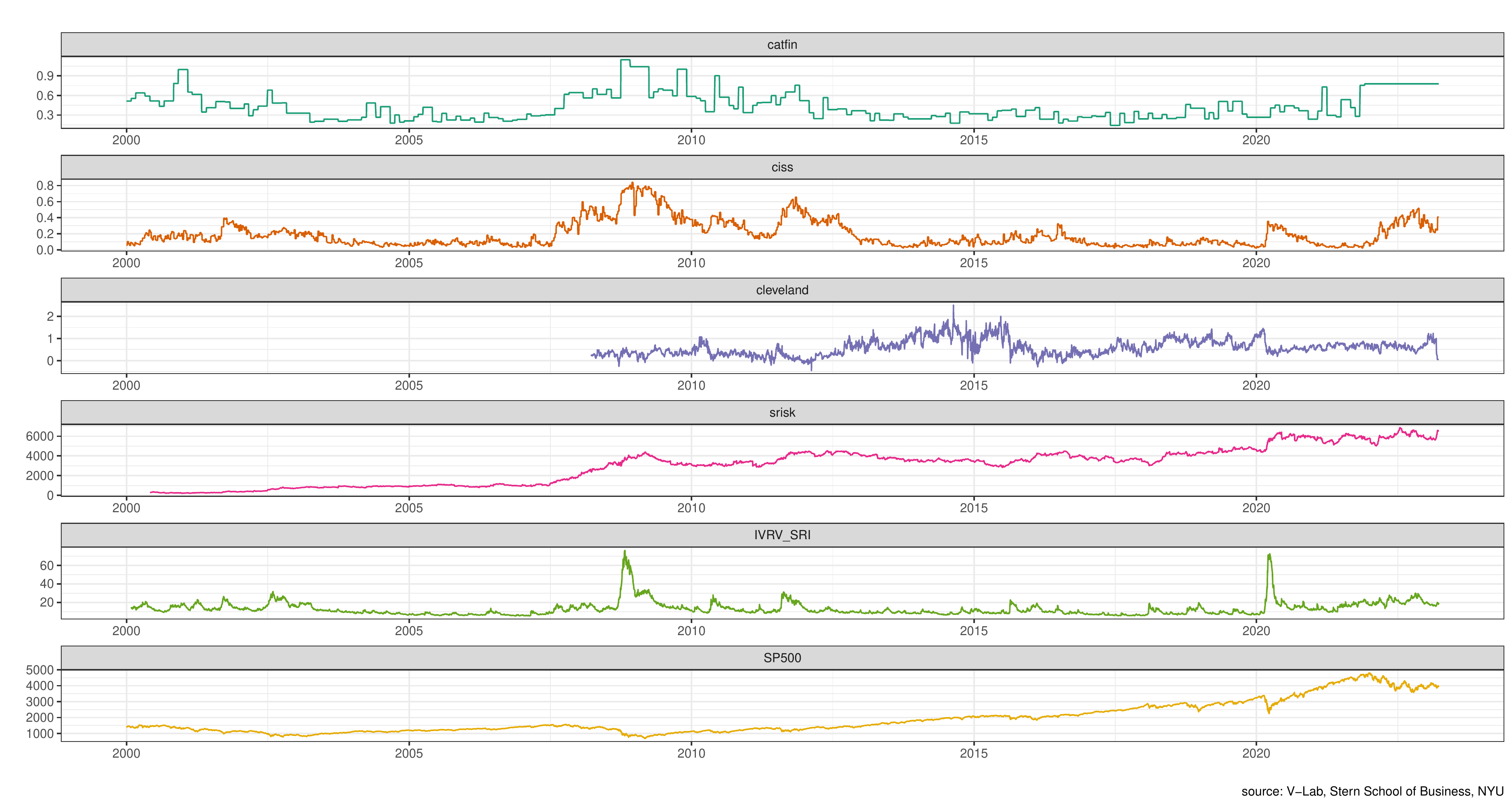}

\caption{Fluctuations of benchmark SRIs and S\&P500 index. \label{fig:fluctuations-levels}}
\scriptsize Note: Figure represents fluctuations of levels of benchmarks systemic risk indicators (SRIs) and SP500 index. Own calculations based on data obtained from V-Lab, Stern School Business, NYU.
\end{figure}

\begin{figure}[H]

\includegraphics[width=1\linewidth]{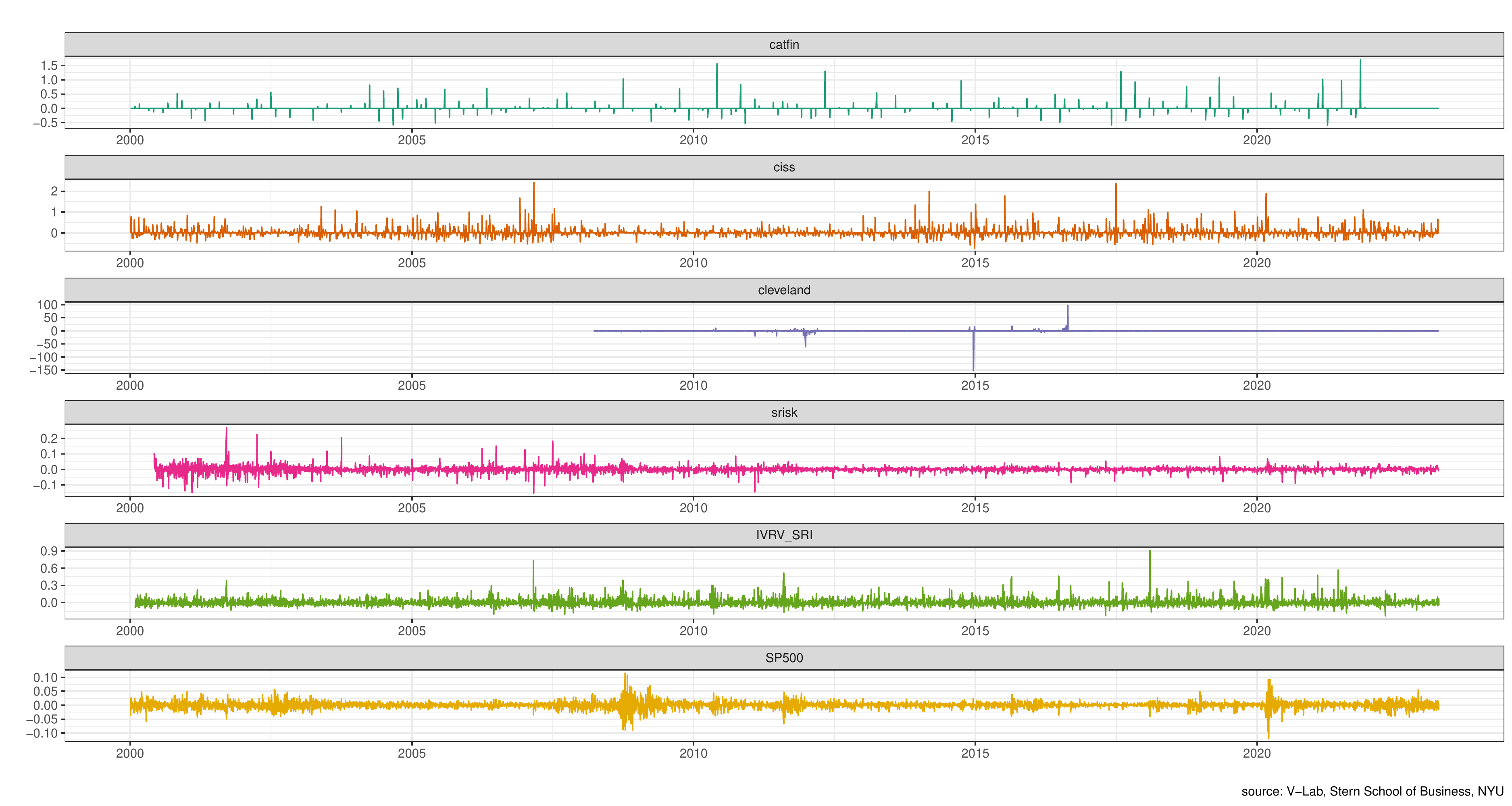}

\caption{Returns of benchmark SRIs, IVRVSRI and S\&P 500 index. \label{fig:fluctuations-returns}}
\scriptsize Note: Figure represents fluctuations of daily simple returns of benchmarks systemic risk indicators (SRIs) and SP500 index. Own calculations based on data obtained from V-Lab, Stern School Business, NYU.
\end{figure}

In this section, in order to see a broader picture for comparison purposes, we present fluctuations of S\&P~500 and analyzed SRIs (Figure \ref{fig:fluctuations-levels}), their weekly returns (Figure \ref{fig:fluctuations-returns}) with descriptive statistics (Table \ref{tab:desc-stats-benchmarks}) and correlations (Table \ref{tab:correlations-returns}), and finally correlation between weekly returns of S\&P 500 index and lagged SRIs (Table \ref{tab:correlations-returns-lagged}).

\begin{table}[H]

\caption{\label{tab:desc-stats-benchmarks}Descriptive statistics of benchmark SRIs, IVRVSRI and S\&P500 index on weekly returns.}
\centering
\fontsize{9}{11}\selectfont
\begin{threeparttable}
\begin{tabular}[t]{>{\raggedright\arraybackslash}p{11em}rrrrrr}
\toprule
statistic & r\_SP500 & r\_catfin & r\_ciss & r\_IVRV\_SRI & r\_srisk & r\_cleveland\\
\midrule
\textbf{nobs} & 5861 & 5861 & 5861 & 5861 & 5861 & 5861\\
\textbf{NAs} & 6 & 5 & 5 & 27 & 111 & 2071\\
\textbf{Minimum} & -0.183 & -0.595 & -0.728 & -0.381 & -0.218 & -673.588\\
\textbf{1. Quartile} & -0.011 & 0 & -0.15 & -0.072 & -0.014 & -0.173\\
\textbf{Mean} & 0.001 & 0.011 & 0.041 & 0.009 & 0.003 & -0.181\\
\textbf{Median} & 0.003 & 0 & -0.002 & -0.013 & 0.002 & -0.001\\
\textbf{3. Quartile} & 0.015 & 0 & 0.154 & 0.059 & 0.019 & 0.185\\
\textbf{Maximum} & 0.191 & 1.7 & 2.423 & 1.82 & 0.317 & 55.463\\
\textbf{Stdev} & 0.025 & 0.161 & 0.316 & 0.144 & 0.037 & 11.262\\
\textbf{Skewness} & -0.56 & 4.173 & 2.102 & 2.981 & 0.962 & -56.49\\
\textbf{Kurtosis} & 6.052 & 34.332 & 9.293 & 20.764 & 8.159 & 3371.311\\
\textbf{Norm.} & 0 & 0 & 0 & 0 & 0 & 0\\
\bottomrule
\end{tabular}
\begin{tablenotes}[para]
\small
\item \footnotesize{Note: Tables represents descriptive statistics of daily simple returns of benchmarks systemic risk indicators (SRIs) and SP500 index. 'Norm.' denotes p-value of the Jarque-Bera test for normality. Own calculations based on data obtained from V-Lab, Stern School Business, NYU.}
\end{tablenotes}
\end{threeparttable}
\end{table}

\begin{table}[H]

\caption{\label{tab:correlations-returns}Correlation matrix for S\&P500 index and SRIs on weekly returns.}
\centering
\fontsize{8}{10}\selectfont
\begin{threeparttable}
\begin{tabular}[t]{>{}l>{\raggedleft\arraybackslash}p{5em}rrrrr}
\toprule
  & r\_catfin & r\_ciss & r\_IVRV\_SRI & r\_cleveland & r\_srisk & r\_SP500\\
\midrule
\textbf{r\_catfin} & 1.000 & 0.012 & 0.066 & 0.002 & 0.062 & -0.054\\
\textbf{r\_ciss} & 0.012 & 1.000 & 0.146 & 0.037 & 0.023 & -0.096\\
\textbf{r\_IVRV\_SRI} & 0.066 & 0.146 & 1.000 & 0.005 & 0.126 & -0.651\\
\textbf{r\_cleveland} & 0.002 & 0.037 & 0.005 & 1.000 & -0.003 & 0.000\\
\textbf{r\_srisk} & 0.062 & 0.023 & 0.126 & -0.003 & 1.000 & -0.228\\
\textbf{r\_SP500} & -0.054 & -0.096 & -0.651 & 0.000 & -0.228 & 1.000\\
\bottomrule
\end{tabular}
\begin{tablenotes}
\small
\item \footnotesize{Note: Table represents correlation matrix for weekly returns of benchmarks systemic risk indicators (SRIs) and SP500 index. Own calculations based on data obtained from V-Lab, Stern School Business, NYU}
\end{tablenotes}
\end{threeparttable}
\end{table}

\begin{table}[H]

\caption{\label{tab:correlations-returns-lagged}Correlation matrix between lagged (5days) weekly returns of SRIs and weekly returns of S\&P500 index.}
\centering
\fontsize{8}{10}\selectfont
\begin{threeparttable}
\begin{tabular}[t]{>{}l>{\raggedleft\arraybackslash}p{5em}rrrrr}
\toprule
  & r\_lag5\_catfin & r\_lag5\_ciss & r\_lag5\_IVRV\_SRI & r\_lag5\_cleveland & r\_lag5\_srisk & r\_SP500\\
\midrule
\textbf{r\_lag5\_catfin} & 1.000 & 0.013 & 0.066 & 0.002 & 0.062 & -0.079\\
\textbf{r\_lag5\_ciss} & 0.013 & 1.000 & 0.147 & 0.038 & 0.022 & -0.028\\
\textbf{r\_lag5\_IVRV\_SRI} & 0.066 & 0.147 & 1.000 & 0.005 & 0.126 & 0.029\\
\textbf{r\_lag5\_cleveland} & 0.002 & 0.038 & 0.005 & 1.000 & -0.003 & 0.018\\
\textbf{r\_lag5\_srisk} & 0.062 & 0.022 & 0.126 & -0.003 & 1.000 & 0.022\\
\textbf{r\_SP500} & -0.079 & -0.028 & 0.029 & 0.018 & 0.022 & 1.000\\
\bottomrule
\end{tabular}
\begin{tablenotes}
\small
\item \footnotesize{Note: Table represents graphical interpretation of correlation matrix for weekly lagged returns of benchmarks systemic risk indicators (SRIs) and weekly returns of SP500 index. Own calculations based on data obtained from V-Lab, Stern School Business, NYU}
\end{tablenotes}
\end{threeparttable}
\end{table}

\hypertarget{forecasting-ability-of-sris}{%
\subsection{Forecasting ability of SRIs}\label{forecasting-ability-of-sris}}

\label{ForecastingAbilityOfSRIs}

In this section we refer to the forecasting ability of IVRVSRI and benchamrk SRIs with regard to the weekly returns of S\&P~500 index. Therefore, we present two set of six models for overlapping and non-overlapping weekly returns. In Tables for simple and quasi-quantile regression we report th adjusted \(R^2\), while in tables for quantile regression the pseudo \(R^2\).

Adjusted \(R^2\) was calculated according to the formula:

\begin{equation}
R^2_{\mathsf{adj}} = 1 - (1 - R^2) \frac{n - 1}{n - p - 1}
\end{equation}
where \(n\) is the number of observations and \(p\) is the number of parameters to estimate.

To calculate pseudo \(R^2\) for the quantile regressions, we follow the \citet{1999koenkermachado} approach, who propose \(R_{\mathsf{pseudo}}^2\) as a local measure of goodness of fit at the particular \(\tau\) quantile. We assume that:
\begin{equation}
V(\tau) = min_b \sum \rho_\tau(y_i - x_i^{'}b)
\end{equation}

Let \(\hat\beta(\tau)\) and \(\tilde\beta(\tau)\) and be the coefficient estimates for the full model, and a restricted model, respectively. Similarly, \(\hat V\) and \(\tilde V\) are the corresponding \(V\) terms. The goodness of fit criterion is then defined as \(R_{\mathsf{pseudo}}^2 = 1 - \hat V / \tilde V\).

The list of tested models is presented in the form two sets of models for overlapping and non-overlapping weekly returns: Simple Regression Models (1-lag and p-lag), Quasi Quantile Regression Models (1-lag and p-lag), and Quantile Regression Models (1-lag and p-lag).

\hypertarget{regression-models-based-on-the-overlapping-data}{%
\subsubsection{Regression models based on the overlapping data}\label{regression-models-based-on-the-overlapping-data}}

The results (Tables \ref{tab:lin-reg-ol}, \ref{tab:qq-reg-ol}, and \ref{tab:q-reg-ol}) of the regressions based on the overlapping data (the simple linear regression, the quasi-quantile regression, and the quantile regression) indicate that the explanatory power of SRIs increase when the larger number of lags is used in the model, and when the regression is performed in a deeper part of the tail of the distribution (quasi-quantile for S\&P500 index returns in the lowest decile and quantile regressions for \(\tau<0.1\)).

However, the most important conclusion from the presented regressions is that the lagged weekly returns of the IVRVSRI have the largest explanatory power of the weekly returns of the S\&P 500 index for all presented regression models.

\begin{table}[H]

\caption{\label{tab:lin-reg-ol}Adjusted R2 for simple regression models based on 1-lag and 
      p-lags for singles SRI and for all SRIs}
\centering
\fontsize{8}{10}\selectfont
\begin{threeparttable}
\begin{tabular}[t]{>{\raggedleft\arraybackslash}p{10em}rrrrrr}
\toprule
\multicolumn{1}{c}{ } & \multicolumn{6}{c}{Adjusted R2} \\
\cmidrule(l{3pt}r{3pt}){2-7}
 & catfin & ciss & IVRV\_SRI & srisk & cleveland & all\\
\midrule
\addlinespace[0em]
\multicolumn{7}{l}{\textbf{lag = 1}}\\
\hspace{1em}\textbf{} & 0.6\% & 0.1\% & 0.1\% & 0.0\% & 0.0\% & 1.0\%\\
\addlinespace[0em]
\multicolumn{7}{l}{\textbf{lag = p}}\\
\hspace{1em}\textbf{} & 2.0\% & 0.7\% & 1.1\% & 0.9\% & -0.4\% & 6.4\%\\
\bottomrule
\end{tabular}
\begin{tablenotes}
\small
\item \footnotesize{Note: Simple regression models based on 1-lag and p-lags were prepared based on Equation \ref{eq:SRegM-1} and Equation \ref{eq:SRegM-2} in the period between 2000 and 2023 on weekly overlapping returns.}
\end{tablenotes}
\end{threeparttable}
\end{table}

\begin{table}[H]

\caption{\label{tab:qq-reg-ol}Adjusted R2 for quasi-quantile regression models based on 1-lag and p-lags for singles SRI and for all SRIs}
\centering
\fontsize{8}{10}\selectfont
\begin{threeparttable}
\begin{tabular}[t]{>{\raggedleft\arraybackslash}p{10em}rrrrr}
\toprule
\multicolumn{1}{c}{ } & \multicolumn{5}{c}{Adjusted R2} \\
\cmidrule(l{3pt}r{3pt}){2-6}
percentile & catfin & ciss & IVRV\_SRI & srisk & cleveland\\
\midrule
\addlinespace[0em]
\multicolumn{6}{l}{\textbf{lag = 1}}\\
\hspace{1em}\textbf{r < mean(r)} & 0.4\% & 1.3\% & 2.5\% & 0.3\% & 0.0\%\\
\hspace{1em}\textbf{r < P25(r)} & 0.5\% & 1.7\% & 4.1\% & 0.2\% & -0.1\%\\
\hspace{1em}\textbf{r < P10(r)} & 1.8\% & 3.6\% & 8.2\% & 0.1\% & -0.2\%\\
\hspace{1em}\textbf{r < P05(r)} & 4.3\% & 4.5\% & 11.3\% & 0.1\% & -0.1\%\\
\hspace{1em}\textbf{r < P01(r)} & 7.7\% & 8.4\% & 16.4\% & -0.7\% & 2.3\%\\
\addlinespace[0em]
\multicolumn{6}{l}{\textbf{lag = p}}\\
\hspace{1em}\textbf{r < mean(r)} & 1.7\% & 3.3\% & 12.5\% & 4.0\% & -0.4\%\\
\hspace{1em}\textbf{r < P25(r)} & 1.0\% & 4.9\% & 14.3\% & 3.0\% & -0.5\%\\
\hspace{1em}\textbf{r < P10(r)} & 0.9\% & 9.9\% & 18.0\% & 4.4\% & 3.6\%\\
\hspace{1em}\textbf{r < P05(r)} & 5.5\% & 16.3\% & 22.1\% & 3.5\% & 6.2\%\\
\hspace{1em}\textbf{r < P01(r)} & 8.3\% & 17.7\% & 20.3\% & -3.3\% & 6.8\%\\
\bottomrule
\end{tabular}
\begin{tablenotes}
\small
\item \footnotesize{Note: Quasi-quantile regression models based on 1-lag and p-lags for overlapping weekly returns were prepared based on Equation \ref{eq:qQRegM-1} in the period between 2000 and 2023 on weekly overlapping returns.}
\end{tablenotes}
\end{threeparttable}
\end{table}

\begin{table}[H]

\caption{\label{tab:q-reg-ol}Pseudo-R2 for quantile regression models based on 1-lag and p-lags for singles SRI and for all SRIs}
\centering
\fontsize{8}{10}\selectfont
\begin{threeparttable}
\begin{tabular}[t]{>{\raggedleft\arraybackslash}p{10em}rrrrr}
\toprule
\multicolumn{1}{c}{ } & \multicolumn{5}{c}{pseudo R2} \\
\cmidrule(l{3pt}r{3pt}){2-6}
tau & catfin & ciss & IVRV\_SRI & srisk & cleveland\\
\midrule
\addlinespace[0em]
\multicolumn{6}{l}{\textbf{lag = 1}}\\
\hspace{1em}\textbf{0.5} & 0.5\% & 0.1\% & 0.6\% & 0.1\% & 0.2\%\\
\hspace{1em}\textbf{0.25} & 0.5\% & 0.1\% & 0.5\% & 0.1\% & 0.2\%\\
\hspace{1em}\textbf{0.1} & 0.5\% & 0.3\% & 0.6\% & 0.1\% & 0.1\%\\
\hspace{1em}\textbf{0.05} & 0.7\% & 0.5\% & 1.2\% & 0.4\% & 0.1\%\\
\hspace{1em}\textbf{0.01} & 4.6\% & 4.0\% & 7.8\% & 1.3\% & 0.1\%\\
\addlinespace[0em]
\multicolumn{6}{l}{\textbf{lag = p}}\\
\hspace{1em}\textbf{0.5} & 4.0\% & 3.0\% & 3.6\% & 2.0\% & 2.9\%\\
\hspace{1em}\textbf{0.25} & 4.5\% & 3.1\% & 4.1\% & 2.6\% & 2.8\%\\
\hspace{1em}\textbf{0.1} & 4.6\% & 3.5\% & 6.9\% & 3.5\% & 2.2\%\\
\hspace{1em}\textbf{0.05} & 4.8\% & 4.7\% & 9.8\% & 4.7\% & 2.2\%\\
\hspace{1em}\textbf{0.01} & 10.0\% & 14.8\% & 23.9\% & 10.2\% & 3.3\%\\
\bottomrule
\end{tabular}
\begin{tablenotes}
\small
\item \footnotesize{Note: Quantile regression models based on 1-lag and $p$-lags for non-overlapping weekly returns were prepared based on Equation \ref{eq:QRegM-1} in the period between 2000 and 2023 on weekly overlapping returns.}
\end{tablenotes}
\end{threeparttable}
\end{table}

\hypertarget{regression-models-based-on-the-non-overlapping-data}{%
\subsubsection{Regression models based on the non-overlapping data}\label{regression-models-based-on-the-non-overlapping-data}}

The results of the regressions based on the non-overlapping data (Tables \ref{tab:lin-reg-nol}, \ref{tab:qq-reg-nol}, and \ref{tab:q-reg-nol}) yield similar conclusions as for the overlapping data. The explanatory power of the lagged weekly returns of the IVRVSRI is still the highest among all analyzed models, measured by the adjusted R2 pseudo-R2 statistics. Moreover, in the case of the IVRVSRI, the predictive power is increasing while analyzing deeper and deeper parts of the tail distributions which is crucial for the point of view of the indicators that measure systemic risk.

\begin{table}[H]

\caption{\label{tab:lin-reg-nol}Adjusted R2 for simple regression models based on 1-lag and p-lags for singles SRI and for all SRIs}
\centering
\fontsize{8}{10}\selectfont
\begin{threeparttable}
\begin{tabular}[t]{>{\raggedleft\arraybackslash}p{10em}rrrrrr}
\toprule
\multicolumn{1}{c}{ } & \multicolumn{6}{c}{Adjusted R2} \\
\cmidrule(l{3pt}r{3pt}){2-7}
 & catfin & ciss & IVRV\_SRI & srisk & cleveland & all\\
\midrule
\addlinespace[0em]
\multicolumn{7}{l}{\textbf{lag = 1}}\\
\hspace{1em}\textbf{} & 0.7\% & -0.1\% & 0.0\% & 0.0\% & -0.1\% & 0.7\%\\
\addlinespace[0em]
\multicolumn{7}{l}{\textbf{lag = p}}\\
\hspace{1em}\textbf{} & 1.4\% & -0.1\% & 0.5\% & 0.5\% & -0.4\% & 2.7\%\\
\bottomrule
\end{tabular}
\begin{tablenotes}
\small
\item \footnotesize{Note: Simple regression models based on 1-lag and p-lags for non-overlapping weekly returns were prepared based on Equation \ref{eq:SRegM-1} and Equation \ref{eq:SRegM-2} in the period between 2000 and 2023 on weekly overlapping returns.}
\end{tablenotes}
\end{threeparttable}
\end{table}

\begin{table}[H]

\caption{\label{tab:qq-reg-nol}Adjusted R2 for quasi-quantile regression models based on 1-lag and p-lags for singles SRI and for all SRIs}
\centering
\fontsize{8}{10}\selectfont
\begin{threeparttable}
\begin{tabular}[t]{>{\raggedleft\arraybackslash}p{10em}rrrrr}
\toprule
\multicolumn{1}{c}{ } & \multicolumn{5}{c}{Adjusted R2} \\
\cmidrule(l{3pt}r{3pt}){2-6}
percentile & catfin & ciss & IVRV\_SRI & srisk & cleveland\\
\midrule
\addlinespace[0em]
\multicolumn{6}{l}{\textbf{lag = 1}}\\
\hspace{1em}\textbf{r < mean(r)} & 1.0\% & 0.6\% & 4.3\% & 0.6\% & -0.3\%\\
\hspace{1em}\textbf{r < P25(r)} & 1.0\% & 0.4\% & 10.3\% & 1.1\% & -0.3\%\\
\hspace{1em}\textbf{r < P10(r)} & 5.7\% & 2.2\% & 24.5\% & 1.5\% & -1.0\%\\
\hspace{1em}\textbf{r < P05(r)} & 10.7\% & 9.7\% & 31.8\% & 2.7\% & 3.3\%\\
\addlinespace[0em]
\multicolumn{6}{l}{\textbf{lag = p}}\\
\hspace{1em}\textbf{r < mean(r)} & 0.2\% & 3.0\% & 15.5\% & 3.4\% & -2.1\%\\
\hspace{1em}\textbf{r < P25(r)} & -0.6\% & 1.5\% & 22.5\% & -0.2\% & 13.7\%\\
\hspace{1em}\textbf{r < P10(r)} & -2.6\% & 10.3\% & 31.8\% & -4.7\% & 9.8\%\\
\hspace{1em}\textbf{r < P05(r)} & -9.6\% & 27.7\% & 40.2\% & 3.1\% & 11.9\%\\
\bottomrule
\end{tabular}
\begin{tablenotes}
\small
\item \footnotesize{Note: Quasi-quantile regression models based on 1-lag and p-lags for non-overlapping weekly returns were prepared based on Equation \ref{eq:qQRegM-1} in the period between 2000 and 2023 on weekly overlapping returns.}
\end{tablenotes}
\end{threeparttable}
\end{table}

\begin{table}[H]

\caption{\label{tab:q-reg-nol}Pseudo-R2 for quantile regression models based on 1-lag and p-lags for singles SRI and for all SRIs}
\centering
\fontsize{8}{10}\selectfont
\begin{threeparttable}
\begin{tabular}[t]{>{\raggedleft\arraybackslash}p{10em}rrrrr}
\toprule
\multicolumn{1}{c}{ } & \multicolumn{5}{c}{pseudo R2} \\
\cmidrule(l{3pt}r{3pt}){2-6}
tau & catfin & ciss & IVRV\_SRI & srisk & cleveland\\
\midrule
\addlinespace[0em]
\multicolumn{6}{l}{\textbf{lag = 1}}\\
\hspace{1em}\textbf{0.5} & 0.5\% & 0.1\% & 1.0\% & 0.1\% & 0.3\%\\
\hspace{1em}\textbf{0.25} & 0.7\% & 0.2\% & 0.8\% & 0.0\% & 0.4\%\\
\hspace{1em}\textbf{0.1} & 0.7\% & 0.3\% & 1.0\% & 0.2\% & 0.2\%\\
\hspace{1em}\textbf{0.05} & 1.3\% & 0.2\% & 2.7\% & 0.9\% & 0.2\%\\
\addlinespace[0em]
\multicolumn{6}{l}{\textbf{lag = p}}\\
\hspace{1em}\textbf{0.01} & 7.8\% & 4.9\% & 14.2\% & 5.5\% & 0.2\%\\
\hspace{1em}\textbf{0.5} & 4.7\% & 3.9\% & 4.5\% & 2.7\% & 4.3\%\\
\hspace{1em}\textbf{0.25} & 4.8\% & 3.9\% & 5.4\% & 3.1\% & 4.7\%\\
\hspace{1em}\textbf{0.1} & 7.0\% & 5.5\% & 10.1\% & 4.3\% & 4.6\%\\
\textbf{0.05} & 9.6\% & 8.4\% & 15.6\% & 6.9\% & 5.6\%\\
\bottomrule
\end{tabular}
\begin{tablenotes}
\small
\item \footnotesize{Note: Quantile regression models based on 1-lag and $p$-lags for non-overlapping weekly returns were prepared based on Equation \ref{eq:QRegM-1} in the period between 2000 and 2023 on weekly overlapping returns.}
\end{tablenotes}
\end{threeparttable}
\end{table}

\hypertarget{conclusions}{%
\section{Conclusions}\label{conclusions}}

\label{Conclusions}

\normalsize

In this study, we propose a robust Systemic Risk Indicator based on the well-known concepts of realized and implied volatility measures. The main contribution of this paper to the broad bulk of studies of systemic risk indicators is the simplicity of the metrics that we propose, which at the same time yield similar results as more complex tools, thus significantly reducing the model risk. At the same time, the proposed methodology enables calculation of IVRVSRI on high-frequency data (even on tick level) which significantly decreases the time of response of our indicator to the starting point of each major financial turmoil. Moreover, in the case of many metrics, it is also much less computationally demanding and does not rely on paid data sets or data that is available only for market regulators. The indication of this measure depends on the geographical location of a given equity market. As expected, the robustness of the proposed Systemic Risk Indicator depends on various parameters selected: the memory parameter for RV, time to expiration for IV, the percentile selected for the risk map, and the length of the history selected for the calculation of percentile in case of the risk map.

Referring to the first hypothesis (RH1), we were able to draw the following conclusions. We can not reject RH1 as we show that it is possible to construct a robust Systemic Risk Indicator (\(\mathrm{IVRVSRI}\)) based on the well-known concepts of realized and implied volatility measures. Moreover, we cannot reject RH2 as the indication of the proposed Systemic Risk Indicator (\(\mathrm{IVRVSRI}_i\)) depends on the geographical location of a given equity market. As expected, the robustness of the proposed Systemic Risk Indicator depends on various parameters selected in the process of its calcuation: the memory parameter for RV, time to expiration for IV, the percentile selected for the risk map, and the length of the history selected for the calculation of percentile in case of risk map, which supports RH3. Finally, the forecasting ability of IVRVSRI is undoubtly the highest from all SRIs used in this study (the srisk, the CATFIN, the Clevelend FED Systemic Risk Indicator, and the CISS). This conclusion is backed by three versions of our regression models (simple, quasi-quantile, quantile) for two types fo weekly returns (overlapping and non-overlapping data).

This study can be extended by adding more countries to the analysis or other asset classes like currencies, commodities, real estate, cryptocurrencies, and hedge funds. Moreover, using high-frequency data would allow the construction of a real-time early implied volatility realized volatility systemic risk indicator (rteIVRVSRI) that would serve as an early warning indicator of systemic risk. We have a plan to design a website within resources of QFRG in order to publish real time data of IVRVSRI and its component parts for theoretical and practical purposes. Finally, there is a need to prepare detailed sensitivity analysis of IVRVSRI for all crucial parameters assumed in the process of its calculations.

\renewcommand\refname{References}
\bibliography{mybibfile.bib}

\end{document}